\documentclass[utf8]{FrontiersinHarvard} 

\usepackage{url,hyperref,lineno,microtype,subcaption}
\usepackage[onehalfspacing]{setspace}
\usepackage{comment}

\newcommand{\chandra}{\textit{Chandra}}

\newcommand{\suzaku}{\textit{Suzaku}}
\newcommand{\xmm}{\textit{XMM}}

\newcommand{\fermi}{\textit{Fermi}}
\newcommand{\swift}{\textit{Swift}}

\newcommand{\nustar}{\textit{NuSTAR}}

\newcommand\hess{{H.E.S.S.\/}}
\newcommand\hexp{\textit{HEX-P}}

\newcommand{\fluxcgs}{\ensuremath{\mathrm{erg}\,\mathrm{s}^{-1}\,\mathrm{cm}^{-2}}}

\def\amin{\ifmmode^{\prime}\else$^{\prime}$\fi}
\def\asec{\ifmmode^{\prime\prime}\else$^{\prime\prime}$\fi}
\def\simgt{\lower.5ex\hbox{$\; \buildrel > \over \sim \;$}}
\def\simlt{\lower.5ex\hbox{$\; \buildrel < \over \sim \;$}}

\def\keyFont{\fontsize{8}{11}\helveticabold }
\def\firstAuthorLast{Mori {et~al.}} 
\def\Authors{Kaya Mori\,$^{1,*}$, Stephen Reynolds\,$^{2}$, Hongjun An\,$^{3}$, Aya Bamba\,$^{4,5,6}$, Roman Krivonos\,$^{30}$, Naomi Tsuji\,$^{7}$,  Moaz Abdelmaguid\,$^{8}$, , Jason Alford\,$^{8}$, Priyadarshini Bangale\,$^{9}$, Silvia Celli\,$^{10}$, Rebecca Diesing\,$^{11}$, Jordan Eagle\,$^{12}$, Chris L. Fryer\,$^{14}$, Stefano Gabici\,$^{15}$, Joseph Gelfand\,$^{8}$, Brian Grefenstette\,$^{16}$, Javier Garcia\,$^{12}$, Chanho Kim\,$^{3}$, Sajan Kumar\,$^{18}$, Ekaterina Kuznetsova\,$^{30}$, Brydyn Mac Intyre\,$^{17}$, Kristin Madsen\,$^{12}$, Silvia Manconi\,$^{19}$, Yugo Motogami\,$^{20}$, Hayato Ohsumi\,$^{20}$, Barbara Olmi\,$^{21,22}$, Jaegeun Park\,$^{3}$, Gabriele Ponti\,$^{23,24}$, Toshiki Sato\,$^{25}$, Ruo-Yu Shang\,$^{26}$, Daniel Stern\,$^{27}$, Yukikatsu Terada\,$^{20,28}$, Jooyun Woo\,$^{1}$, George Younes\,$^{12,13}$, and Andreas Zoglauer\,$^{29}$}
\def\Address{\footnotesize $^{1}$ Columbia Astrophysics Laboratory, Columbia University, New York, NY, USA\\
$^{2}$ Department of Physics, North Carolina State University, Raleigh, NC, USA\\  
$^{3}$ Department of Astronomy and Space Science, Chungbuk National University, Cheongju, Republic of Korea\\  
$^{4}$ Department of Physics, The University of Tokyo, Tokyo, Japan\\  
$^{5}$ Research Center for the Early Universe, School of Science, The University of Tokyo, Tokyo, Japan\\
$^{6}$ Trans-Scale Quantum Science Institute, The University of Tokyo, Tokyo, Japan\\
$^{7}$ Faculty of Science, Kanagawa University, Kanagawa, Japan\\  
$^{8}$ Department of Physics, New York University, Abu Dhabi, UAE\\  
$^{9}$ Department of Physics and Astronomy and the Bartol Research Institute, University of Delaware, Newark, DE, USA\\
$^{10}$ Department of Physics, Sapienza University of Rome and INFN Sezione di Roma Piazzale Aldo Moro, Rome, Italy\\  
$^{11}$ Department of Astronomy and Astrophysics, The University of Chicago, Chicago, IL, USA\\  
$^{12}$  NASA Goddard Space Flight Center, Greenbelt, MD, USA\\  
$^{13}$ Department of Physics, The George Washington University, Washington, DC, USA\\  
$^{14}$ Center for Non Linear Studies, Los Alamos National Laboratory, Los Alamos, NM, USA\\  
$^{15}$ Universite Paris Cite, CNRS, Astroparticule et Cosmologie, Paris, France\\  
$^{16}$ Space Radiation Laboratory, California Institute of Technology,  Pasadena, CA USA\\  
$^{17}$ Department of Physics and Astronomy, University of Manitoba, Winnipeg, Canada\\  
$^{18}$ Department of Physics, University of Maryland, College Park, MD 20742-4111, USA\\ 
$^{19}$ Laboratoire d’Annecy-le-Vieux de Physique Theorique, CNRS, USMB, Annecy, France\\ 
$^{20}$ Graduate School of Science and Engineering, Saitama University, Sakura, Saitama, Japan\\ 
$^{21}$ INAF - Osservatorio Astrofisico di Arcetri, Largo E. Fermi 5, I-50125 Firenze, Italy\\
$^{22}$ INAF - Osservatorio Astronomico di Palermo, Piazza del Parlamento 1, I-90134 Palermo, Italy\\
$^{23}$ INAF - Osservatorio Astronomico di Brera, Merate, Italy \\ 
$^{24}$ Max-Planck-Institut für extraterrestrische Physik, Garching, Germany \\
$^{25}$ Department of Physics, School of Science and Technology, Meiji University, Kanagawa, Japan\\  
$^{26}$ Department of Physics and Astronomy, Barnard College, New York, NY, USA\\  
$^{27}$ Jet Propulsion Laboratory, California Institute of Technology, Pasadena, CA, USA\\
$^{28}$ Japan Aerospace Exploration Agency(JAXA), Institute of Space and Astronautical Science, Sagamihara, Japan\\ 
$^{29}$ Space Sciences Laboratory, UC Berkeley, Berkeley, CA, USA\\  
$^{30}$ Space Research Institute (IKI), Russian Academy of Sciences, Moscow 117997, Russia
}
\def\corrAuthor{\footnotesize Kaya Mori}
\def\corrEmail{\footnotesize kaya@astro.columbia.edu}

\begin{document}
\onecolumn
\firstpage{1}

\title {The High Energy X-ray Probe (HEX-P): Galactic PeVatrons, star clusters, superbubbles, microquasar jets, and gamma-ray binaries} 

\author[\firstAuthorLast ]{\Authors} 
\address{} 
\correspondance{} 

\extraAuth{}

\maketitle

\begin{abstract}

\section{}

HEX-P is a probe-class mission concept that will combine high spatial resolution X-ray imaging ($<10''$ FWHM) and broad spectral coverage (0.2--80 keV) with an effective area far superior to current facilities (including XMM-Newton and NuSTAR) to enable revolutionary new insights into a variety of important astrophysical problems. With the recent discoveries of over 40 ultra-high-energy gamma-ray sources (detected above 100 TeV) and neutrino emission in the Galactic Plane, we have entered a new era of multi-messenger astrophysics facing the exciting reality of Galactic PeVatrons. In the next decade, as more Galactic PeVatrons and TeV gamma-ray sources are expected to be discovered, the identification of their acceleration and emission mechanisms  will be the most pressing issue in both particle and high-energy astrophysics.   
In this paper, along with its companion papers (Reynolds et al. 2023, Mori et al. 2023), we will 
present that HEX-P is uniquely suited to address important problems in various cosmic-ray accelerators, including Galactic PeVatrons, through investigating synchrotron X-ray emission of TeV--PeV electrons produced by both leptonic and hadronic processes. 
For Galactic PeVatron candidates and other TeV gamma-ray sources, HEX-P can fill in a large gap in the spectral-energy distributions (SEDs) of many objects observed in radio, soft X-rays, and gamma rays, constraining the maximum energies to which electrons can be accelerated, with implications for the nature of the Galactic PeVatrons and their contributions to the spectrum of Galactic cosmic rays beyond the knee at $\sim3$ PeV. In particular, X-ray observation with HEX-P and TeV observation with CTAO will provide the most powerful multi-messenger diagnostics to identify Galactic PeVatrons and explore a variety of astrophysical shock mechanisms.  
We present simulations of each class of Galactic TeV--PeV sources, demonstrating the power of both the imaging and spectral capabilities of HEX-P to advance our knowledge of Galactic cosmic-ray accelerators. In addition, we discuss HEX-P’s unique and complementary roles to upcoming gamma-ray and neutrino observatories in the 2030s. 


\tiny
 \keyFont{ \section{Keywords:} particle accelerators, Galactic PeVatrons, star clusters, superbubbles, microquasars, gamma-ray binaries, X-ray telescopes, multimessenger astronomy} 
\end{abstract}

\section{Introduction}

Over the last few decades, it has become clear that energetic
particles (cosmic rays, CRs) make up a significant component of the Universe.  In
galaxies, cosmic rays can power galactic winds, support galactic
coronae, and control star formation through ionization of molecular
clouds.  They can influence the structure of large-scale galactic
magnetic fields and their propagation can drive turbulence in the
interstellar medium. (See, e.g., \cite{2022ApJ...941...78H} and
references therein.)  The most energetic cosmic rays, with energies
from $10^{15}$ eV (1 PeV) up to and above $10^{19}$ eV, appear to fill
the Universe, traveling enormous distances to arrive at Earth -- the
only form of extragalactic matter we will be able to directly examine.

X-ray astronomy has brought powerful insights into the mechanisms by
which Nature accelerates particles to energies many orders of
magnitude above thermal energies.  While radio astronomy even from its
infancy gave evidence of electrons with GeV energies (through the
diffuse Galactic synchrotron background discovered by chance by Karl
Jansky in 1932, though its origin wasn't clear for decades), it was
known long before the advent of space astronomy that far higher
energies were exhibited by some particles, to the extent that
high-energy physics experiments were conducted on mountaintops to tap
the flux of incoming cosmic rays, long before the advent of terrestrial particle
accelerators. But clues to the origin of the highest-energy particles required the ability to image sources in photons above the optical window.  Quasar
continua supplied evidence for optical synchrotron radiation, but the
details of the process could not be deduced from unresolved
observations.  X-ray astronomy first allowed the inference of the
presence of TeV particles at their sources, with the detection of the
featureless X-ray spectrum of SN 1006 and its interpretation as
synchrotron emission from electrons with such energies.

The advent of diffusive shock acceleration (DSA) as a mechanism for
the production of suprathermal particles in shocks constituted a major
advance in understanding, along with the observational data supplied
by several generations of X-ray satellites, most importantly \chandra\ 
and \xmm.  It has been well-established that TeV electrons are
present in most young shell supernova remnants (SNRs), with about ten
objects dominated by non-thermal synchrotron emission, and clear
non-thermal spectral components in others alongside thermal emission.  The initial
hope that young SNRs could furnish the origin of all cosmic rays was
dashed by the realization based on very general considerations that
standard SNR evolution could produce energies only up to several PeV, 
where a steepening of the integrated cosmic ray spectrum suggests a 
decreasing efficiency of cosmic-ray production by Galactic sources. 
More detailed study strongly suggests that reaching even that energy may be difficult \citep[e.g.,][]{lagage83}. See \cite{blasi13} for a review.

More recently, gamma-ray astronomy has revealed many Galactic sources
with energies in the GeV range (observed by \fermi-LAT) to above 1 PeV
(observed by ground-based imaging atmospheric Cerenkov telescopes (IACTs), such as VERITAS, MAGIC, and H.E.S.S., or extensive air-shower arrays (EASAs), such as HAWC and LHAASO). Recently, a new exciting discovery has been made by IceCube as they have identified neutrino emission in the Galactic Plane \citep{icecube2023}. 
Angular resolutions of these instruments often do not allow
unambiguous identification with sources at lower photon energies, let
alone provide morphological clues to the origins of the fast particles
in those sources.  The imaging capabilities of X-ray telescopes,
current and planned, can address the gaps in our understanding
resulting from the mismatch between high-resolution radio observations
of GeV electrons, and the observations of particles (electrons or
hadrons) of up to and above 1 PeV.

Various classes of objects are now known to produce energetic
particles: SNRs, pulsar-wind nebulae (PWNe) at the termination shock
of the pulsar's relativistic wind forming the inner boundary of the
PWN, superbubbles driven by multiple supernovae, and termination
shocks of jets from "microquasars" such as SS433.  Arguments have been
made for each of these as the primary source of the most energetic
Galactic cosmic rays, but in no case do we have conclusive
determinations.

A full understanding of the physics of particle acceleration in
shocks, and elsewhere, and in particular, of the nature of the most
energetic sources (``PeVatrons"), will require a new generation of
instruments.  In the context of DSA, we still do not understand the
details of how electrons become initially accelerated; how the
accelerated-particle population, both electrons and ions, develops and affects the local environment (magnetic field, thermal fluid); what determines the fractions of shock energy going into particles and magnetic field; and what determines the maximum energy to which particles are accelerated, with possibly different limitations applying to electrons and hadrons. For the wind termination shocks of PWNe, the additional complications of special-relativistic effects are present.

In particular, an attack on the problem of the nature of ``PeVatrons"
can be divided into two fronts: a better understanding of the basic
physics of particle acceleration, conducted in those objects which can
be most fully characterized, and direct observational studies of 
candidate PeVatrons themselves.  The latter project is hampered by the
large point-spread functions (PSFs) of Cherenkov detectors (of order a significant fraction of a
degree), often containing multiple possible counterparts in the
crowded Galactic plane.  In the former approach, one works to 
improve our knowledge of the basic physics of shock acceleration by studying better-understood objects, but ones fairly certain not to be the PeVatrons themselves.  The proposed HEX-P mission can contribute
on both fronts. 

In this paper, we present a wealth of HEX-P programs for investigating a diverse class of cosmic-ray accelerators and exotic radioactive sources in our Galaxy. In \S2, we review the key radiative processes as a primer for understanding multi-wavelength electromagnetic emission from cosmic-ray accelerators. \S3 and \S4 describe the current telescope design and HEX-P's primary observation program for Galactic cosmic-ray accelerators, respectively. The HEX-P's primary observation program has been optimally determined based on extensive simulations with the Simulations of X-ray Telescopes (SIXTE) suite and with NASA's HEASARC XSPEC  software, as well as consulting with the current and future gamma-ray and neutrino telescope groups, including CTAO, HAWC, VERITAS, IceCube, and COSI. Note that two primary classes of Galactic particle accelerators, SNRs and PWNe, are discussed in a companion paper (Reynolds et al. 2023). 
\S5 discusses the unique and complementary role of HEX-P in the future multi-messenger observations of Galactic PeVatrons, which is arguably the most exciting field in astroparticle physics currently and in the 2030s. 
\S6 presents HEX-P observations of star clusters and superbubbles which represent a primary class of hadronic particle accelerators. \S7 focuses on the HEX-P survey of W50 lobes, a unique particle accelerator powered by the microquasar SS433. \S8  presents how HEX-P observations can deepen our understanding of  intrabinary shock physics and interactions between pulsars and circumstellar disks in rare TeV gamma-ray binaries. \S9  concludes the paper with various HEX-P survey ideas in synergy with future telescopes in other wavelengths.


\section{Radiative Processes}

Accelerated particles make their presence known through a variety of
radiative processes: synchrotron radiation, bremsstrahlung, and
inverse-Compton scattering for the electrons or positrons (leptonic
processes), and, for protons and nuclei, decay into gamma rays of
$\pi^0$ mesons produced in inelastic scattering from target atoms
(hadronic process).

Relativistic electrons (or positrons) of energy $E$ radiating in a
magnetic field {\bf B} produce synchrotron radiation with a spectrum
peaking at photon energy $E_{\gamma} = 15 \langle B_{\mu{\rm
G}}\rangle \, E_{\rm PeV}^2$ keV, after averaging over the angle between
{\bf B} and the line of sight.  These particles can also upscatter any
ambient photon fields through inverse-Compton scattering (ICS).  Both
cosmic microwave background (CMB) photons and optical-IR (OIR) photons
can serve as relevant seed photon populations.  The scattering cross-section is constant at its Thomson value $\sigma_T \equiv
6.65 \times 10^{-25}$ cm$^{2}$ for small values of the Klein-Nishina
(KN) parameter $x_{\rm KN} \equiv 4E\, E_{\gamma i}/(m_e c^2)^2$,
where $E_{\gamma i}$ is the seed photon energy.  But as $x_{\rm KN}$
approaches and exceeds 1, the cross-section decreases (Klein-Nishina
suppression).  The maximum outgoing photon energy $E_\gamma$ is given,
for $x_{\rm KN} \ll 1$, by $E_{\gamma} = 4 (E/m_e c^2)^2 E_{\gamma i}
= x_{\rm KN} E$, but as $x_{\rm KN}$ approaches 1, $E_{\gamma}$
asymptotes to $E$, independent of the seed photon energy.  For CMB
seeds ($E_{\gamma i} \sim 0.2$ meV), requiring $x_{\rm KN} \leq 0.1$ to
remain safely in the Thomson limit, the peak scattered photon energy
is about 3 TeV, produced by electrons with $E \sim 30$ TeV.
Higher-energy photons can be produced, but with decreasing efficiency.
For OIR seeds ($E_{\gamma i} \sim 1$ eV), again requiring $x_{\rm KN}
< 0.1$ limits scattered photon energies to about 400 MeV, produced by
electrons with $E \sim 7$ GeV.  Thus ICS from the CMB can produce very high-energy (VHE, 100 GeV -- 100 TeV) 
gamma-rays detectable with IACTs, while ICS from starlight seeds is
most important below 1 GeV photon energies, observable with
satellites.  Finally, relativistic electrons can also produce
relativistic bremsstrahlung with photon energies up to $E_\gamma \sim
E/3$, but for the regions of parameter space relevant to the particle
acceleration sources in this paper, bremsstrahlung is rarely dominant.

Cosmic-ray protons and nuclei can produce gamma-ray emission through
inelastic collisions with ambient gas, resulting in the production of
pions (hadronic process).  The charged pions decay to secondary
electrons and positrons, while the $\pi^0$ particles decay to
gamma rays.  These collisions result in a fixed ratio of gamma rays to
secondary leptons, an unavoidable consequence of the process.  For
kinematic reasons, pions cannot be produced until proton energies
reach 280 MeV, at which point it becomes possible to produce
$\pi^0$'s. 

The synchrotron process does not contribute to gamma-ray emission, but
it plays a crucial role in providing evidence for the highest-energy
electrons.  Electrons with energies above a few tens of TeV scatter
CMB photons much less efficiently, so their gamma-ray emission may be
faint or lost below hadronic gamma-ray processes.  However, their
maximum energy can
be constrained through their synchrotron radiation, in the energy range
targeted by HEX-P.

Thus a population of relativistic leptons and hadrons produces
broadband emission from radio to above PeV energies, through all four
processes in general.  Characterizing those populations requires
observations at all wavelengths, to create a spectral-energy distribution (SED).  Radio emission, produced only by
$\sim$ GeV-range electrons, can anchor the lepton distribution.  The
gamma-ray part of the spectrum above 70 MeV could be produced by two
leptonic and one hadronic process, and sorting out which is
responsible is an essential task for the investigation of particle
acceleration.  Of particular interest are instruments capable of
observing very high-energy and ultra high-energy (UHE, $> 100$ TeV) gamma rays.  While most of these instruments
have angular resolutions of a significant fraction of a degree, 
the Cherenkov Telescope Array (CTA), currently under construction and
scheduled to begin full operation in the mid-2020s, is of particular
interest for the HEX-P mission.  Two CTAO sites in the northern and southern hemispheres will be able to survey the entire sky in the 0.1--100 TeV band with $\sim1$ arcmin angular resolution.

\begin{figure}[ht!]
    \centering
    \includegraphics[width=1.0\linewidth]{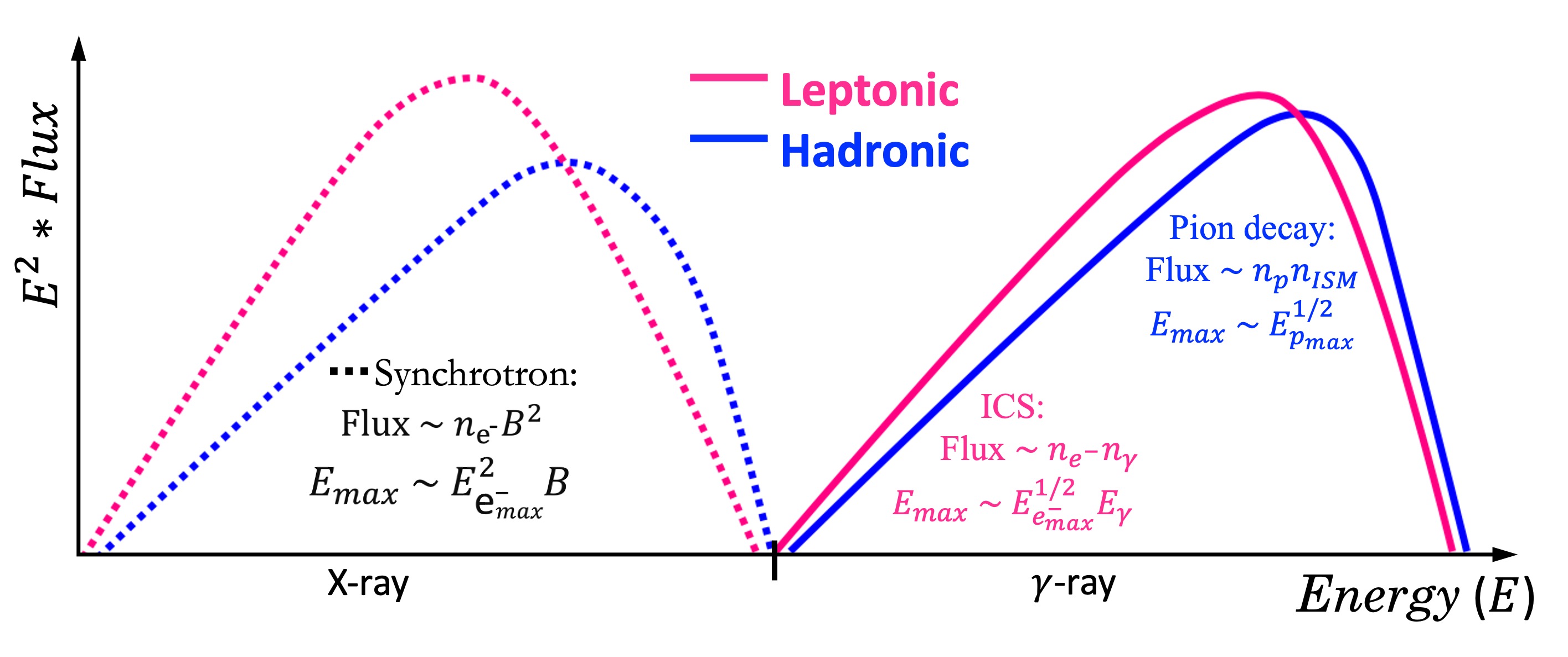}
    \caption{Left: Cartoon multiwavelength (MW) SED models for leptonic and hadronic accelerators. Right: Example leptonic (black) and hadronic (red) SED models for a hypothetical gamma-ray emitting particle accelerator. The two SED models with nearly identical gamma-ray spectra were produced using {\tt NAIMA} \citep{naima}. Note that the synchrotron X-ray spectra appear differently in both fluxes and slopes in the HEX-P bandpass (0.2--80 keV). The secondary electrons in the hadronic case are modeled following the recipe of \citet{Kelner2006}.   
    }
    \label{fig:pevatron_sed}
\end{figure}

In general, synchrotron X-ray emission ($F_X \propto n_e \cdot  B^2$), ICS gamma-ray emission ($F_\gamma \propto n_e \cdot  n_\gamma$), and hadronic gamma-ray emission ($F_\gamma \propto n_p \cdot  n_{\rm ISM}$), where $n_e$, $n_\gamma$, $n_p$ and $n_{\rm ISM}$ are electron, seed photon, proton, and ISM densities, represent different components from the same underlying particle energy distribution (Figure \ref{fig:pevatron_sed}). Hence, in numerous Galactic TeV sources, 
multiwavelength morphology and SED studies with X-ray and IACT TeV data helped to distinguish between the leptonic and hadronic scenarios and to constrain model parameters  \citep{Kargaltsev2013, Mori2021}. For instance, Figure \ref{fig:pevatron_sed} illustrates  the distinct X-ray spectra expected in leptonic and hadronic models, despite predicting nearly identical gamma-ray spectra.  Additional constraints come from the fact that the same
population of relativistic electrons can produce both synchrotron and
inverse-Compton emission.  The ratio of total power radiated by an electron in the two processes is given by the ratio of magnetic-field energy density to seed photon energy density, i.e., $P_{\rm synch}/P_{\rm IC} = [(B^2/8\pi)/u_{\rm rad}]$.  The radiated spectra have the same slope up to peak energies related to the maximum electron energy.  
The synchrotron peak depends on the magnetic field,
while the IC peak depends on which seed photons are scattered.
Therefore, the ratio of the peak photon energies can give an estimate
of the magnetic-field strength, while the ratio of the peak fluxes there
can give the magnetic-field filling factor (i.e., if the volume of
radiating electrons is not uniformly filled with $B$).  See \cite{aharonian99} and \cite{lazendic04} for detailed expressions.

While gamma-ray detections provide evidence of particle acceleration, studying synchrotron X-ray emission from primary and secondary electrons  provides unique and complementary information to gamma-ray observations. However, X-ray telescopes operating only below 10 keV are hampered by the contamination by unrelated  thermal X-ray components, which can hinder the detection of non-thermal X-ray emission. 

\section{HEX-P mission design and simulation}

The High-Energy X-ray Probe (HEX-P; Madsen+23) is a probe-class mission concept that offers sensitive broad-band coverage ($0.2-80$\,keV) of the X-ray spectrum with exceptional spectral, timing, and angular capabilities. It features two high-energy telescopes (HET) that focus hard X-rays, and a low-energy telescope (LET) providing soft X-ray coverage. Overall, LET and HET will achieve a factor of $\sim3$ and $\sim10$ or better improvements in sensitivity over \xmm-EPIC and \nustar\ telescopes in the 0.2-20 and 3-70 keV bands, respectively.

The LET consists of a segmented mirror assembly coated with Ir on monocrystalline silicon that achieves a half power diameter of $3.5''$, and a low-energy DEPFET detector, of the same type as the Wide Field Imager (WFI; Meidinger et al. 2020) onboard Athena (Nandra et al., 2013). It has 512 x 512 pixels that cover a field of view of $11.3' \times 11.3'$. It has an effective bandpass of $0.2-25$\,keV and a full-frame readout time of 2\,ms, and can be operated in a 128 and 64 channel window mode for higher count-rates to mitigate pile-up and faster readout. Pile-up effects remain below an acceptable limit of $\sim 1$\% for sources up to $\sim 100$\,mCrab in the smallest window configuration (64w). Excising the core of the PSF, a common practice in X-ray astronomy, will allow for observations of brighter sources, with a maximum loss of $\sim 60\%$ of the total photon counts.

The HET consists of two co-aligned telescopes and detector modules. The optics are made of Ni-electroformed full shell mirror substrates, leveraging the heritage of \xmm\  \citep{Jansen2001}, and coated with Pt/C and W/Si multilayers for an effective bandpass of $2-80$\,keV. The high-energy detectors are of the same type as those flown on \nustar\ \citep{Harrison2013}, and they consist of 16 CZT sensors per focal plane, tiled $4 \times 4$, for a total of $128 \times 128$ pixel spanning a field of view slightly larger than for the LET, of $13.4' \times 13.4'$.

All the simulations presented here were produced with a set of response files that represent the observatory performance based on current best estimates (see Madsen+23). The effective area is derived from a ray-trace of the mirror design including obscuration by all known structures. The detector responses are based on simulations performed by the respective hardware groups, with an optical blocking filter for the LET and a Be window and thermal insulation for the HET. The LET background was derived from a GEANT4 simulation \citep{Eraerds2021} of the WFI instrument, and the one for the HET from a GEANT4 simulation of the \nustar\ instrument, both positioned at L1. Throughout the paper, we present our simulation results for \hexp\ using the SIXTE \citep{Dauser2019} and XSPEC toolkits \citep{Arnaud1996}. To ensure the most realistic simulation results, we incorporated recent high-resolution X-ray images (mostly from \chandra\ or other wavelength observations), the best-known  spectral information, and theoretical model predictions. Various exposure times have been considered for the feasibility studies presented in the following sections.

\section{HEX-P observation program overview}



One of the main objectives for the HEX-P mission is to comprehensively explore all types of cosmic-ray accelerators in our Galaxy, including the recently discovered PeVatron candidates. HEX-P's primary mission will include observations of a diverse class of Galactic particle accelerators as outlined in Table \ref{tab:obs}. Broadly speaking, understanding particle acceleration, propagation, and cooling entails tackling four-dimensional problems that involve spatial distribution $(X, Y)$, particle energy ($E$), and time ($t$). Modeling and observing particle acceleration/injection sites serve as the initial step for this exploration. While pulsars act as single-point central engines for producing PWNe, SNRs exhibit multiple acceleration sites, such as forward and reverse shock waves. 
Once particles are accelerated and injected into the ambient medium, they propagate and lose kinetic energy through various mechanisms, including radiative loss, adiabatic cooling, and collisions with the ISM and molecular clouds. Different regions away from the central engines contain particles injected at different times, requiring multi-zone investigations to track particle transport and cooling. Resolving the multi-wavelength  radiation from multiple regions is vital to grasp the entire picture of how particle acceleration, propagation, and cooling interplay with each other. Furthermore, to assess the contribution of specific types of particle accelerators to the local and global CR populations in our Galaxy, it is crucial to examine objects at different stages of evolution.

Typically, energetic and powerful particle accelerators have been discovered through gamma-ray observations.  Since the directional and intrinsic energy information of CRs is lost by the interstellar magnetic fields, the CR acceleration and propagation can only be probed indirectly through associated TeV gamma-ray sources. While gamma-ray detections indicate the presence of particle acceleration, the limited angular resolution of TeV gamma-ray telescopes often prevented source identifications. Multi-wavelength observations are required to determine source types and acceleration mechanisms, utilizing telescopes at lower energies and with sub-arcminute angular resolutions. 

While CTAO is expected to revolutionize our views of Galactic particle accelerators in the TeV band, its high-quality gamma-ray data alone cannot fully identify the source types and acceleration mechanisms since broader spectral data are required to separate out different potential emission components (ICS, pion-decay, and synchrotron radiation). It has long been realized that multi-wavelength observations are crucial for identifying the sources and elucidating their emission/acceleration mechanisms. At lower energies, X-ray and radio observations play a unique and complementary role to the gamma-ray data by detecting synchrotron radiation from GeV--PeV electrons. Furthermore, broad-band X-ray spectral data provide unique diagnostics for determining the highest energy cutoff region of hadronic PeVatrons \citep{Celli2020}.  To do this, the detailed shape of the X-ray spectrum, in particular
the existence of curvature or spectral breaks must be well characterized.

Above all, the necessary condition for studying any particle accelerator in the X-ray band is the clean detection of non-thermal X-ray emission apart from soft, thermal X-rays. While previous and current X-ray telescopes such as \nustar\ achieved some success in this respect, HEX-P surpasses them as the ultimate non-thermal X-ray detector, given its unprecedented sensitivity above 10 keV. HEX-P excels in spatially resolving thermal and non-thermal (synchrotron) X-ray emission, providing valuable broadband X-ray morphology and spectroscopy data. Moreover, determining the distribution of magnetic fields around the acceleration sites is important for understanding particle transport and cooling processes. For example, low ambient B-fields may offer insights into why some PeVatron candidates seem to sustain extended TeV emission  without undergoing fast synchrotron cooling. Variability studies with HEX-P, as well as modeling small-scale features, can constrain magnetic field strengths.
HEX-P alone or in synergy with future TeV telescopes, provides the most powerful diagnostic tools to investigate the creation and propagation of the most energetic particles as well as their environments (e.g., magnetic field), ultimately shaping the CR populations both below and above the knee at $\sim 3$ PeV. A golden combination of HEX-P and CTAO can usher in a new and exciting era of multi-zone and multi-wavelength approaches to studying Galactic particle accelerators. 

Below we summarize the primary HEX-P programs for studying a diverse class of Galactic particle accelerators as well as for investigating
nucleosynthesis in young SNRs and potential neutron-star mergers.  More detailed descriptions and simulation results on each program can be found in the subsequent sections. 

\textbf{(1) Investigation of a variety of astrophysical shocks in the primary CR accelerators} through synchrotron X-ray emission of TeV--PeV electrons: (1) DSA in SNRs, (2) PWN termination shock, (3) interactions between the SNR reverse shock and PWN, (4) intra-binary shock in gamma-ray binaries, (5) microquasar jet-driven shock, (6) colliding wind shock in star clusters, and (7) supermassive blackhole at Sgr A*. As outlined in Table \ref{tab:obs}, the primary science program will observe all these types of particle accelerators and prompt follow-up HEX-P observations. 

\textbf{(2) Broad-band X-ray census of Galactic PeVatron candidates:} In the most exciting and unexplored regime of astroparticle physics, HEX-P will play a pivotal role in multi-wavelength astrophysics by identifying Galactic PeVatrons and their acceleration/emission mechanisms in synergy with the upcoming CTAO observatory. Currently, 43 PeVatron candidates have been detected by LHAASO \citep{lhasso23}, but most of them have not yet been identified with
known sources, and even the leptonic or hadronic nature of the emission
is not known.  By the 2030s, more PeVatron candidates (at least $\sim100$) are expected to be discovered. In addition, measuring the maximum particle energy exceeding $\sim1$ PeV is necessary to establish that the LHAASO sources are indeed PeVatrons. HEX-P and CTAO will determine both the accelerator types and maximum particle energies through multi-wavelength SED studies. 

\textbf{(3) Providing a dynamic view of X-ray filaments and knots in young SNRs: } HEX-P will be able to detect year-scale variability from X-ray knots in young SNRs due to ongoing particle acceleration, magnetic field amplification, and fast synchrotron cooling (e.g., $\tau < 6$ yr for  electrons emitting synchrotron X-rays at $E_\gamma > 40$ keV and $B=0.1$ mG). HEX-P will identify the most energetic acceleration sites and determine if young SNRs contain localized PeVatrons associated with hard X-ray knots. This investigation is uniquely conducted with HEX-P by measuring local B-fields and maximum electron energies. See more details in the HEX-P SNR/PWN paper (Reynolds et al. 2023).

\textbf{(4) Dissecting particle acceleration, propagation, and cooling mechanisms in PWNe: } HEX-P can provide spatially-resolved X-ray spectroscopy and broad-band X-ray morphology data of young and evolved PWNe. The pulsar X-ray emission can be separated and studied by timing analysis thanks to HEX-P's $< 2$ ms temporal  resolution. HEX-P alone enables comprehensive multi-zone investigations to understand how particle acceleration, propagation, and cooling operate across various stages of PWN evolution. See more details in the HEX-P SNR/PWN paper (Reynolds et al. 2023).

\textbf{(5) Searching for non-thermal X-ray emission from star clusters and superbubbles: } HEX-P will search for non-thermal X-ray emission arising from colliding winds in the prominent star clusters and superbubbles such as Arches, Westerlund 1 \& 2, Cygnus OB region and 30 Dor C. These star clusters are recognized as primary sites for hadronic acceleration and are largely responsible for producing extended gamma-ray cocoons.

\textbf{(6) Surveying particle accelerators and CR distributions in the Galactic Center (GC): } In the GC, HEX-P will conduct surveys of the primary accelerators, including the supermassive BH at Sgr A*, the youngest SNR in our Galaxy (G1.9+0.3) and  bright TeV sources such as PWN G0.9+0.1. HEX-P will also map the spatial and energy distributions of CRs in the central molecular zone and through X-ray filaments in the GC. See more details on Sgr A* flares, X-ray filaments and Arches cluster in the HEX-P GC paper (Mori et al. 2023).


\begin{table*}
\renewcommand\thetable{1}
{\small
\begin{center}
\caption{HEX-P primary observation program of Galactic cosmic-ray accelerators}
\begin{tabular}{lcccc}
\hline\hline
Source name & Source type & Exposure time (ks) & Section  \\ 
\hline\hline
LHAASO sources & Unidentified PeVatrons & $5\times200$ ks  & \S5 \\ 
Cas A & Young SNR & 200 ks & SNR/PWN paper (\S3.3)\\
Tycho  & Young SNR & 200 ks & SNR/PWN paper (\S3.4) \\
G1.9+0.3 & Young SNR & 200 ks & SNR/PWN paper (\S3.5)\\  
SN1987A & Young SNR + PWN & 300 ks & SNR/PWN paper (\S3.6) \\
Crab  & Young PWN & Calibration source & SNR/PWN paper (\S4.3) \\ 
G21.5-0.9  & Young PWN &  Calibration source & SNR/PWN paper (\S4.3) \\
Lighthouse nebula & Middle-aged PWN & 100 ks & SNR/PWN paper (\S4.3) \\
G0.9+0.1 & Young PWN & 100 ks & SNR/PWN paper (\S4.3)\\ 
Arches cluster & Star cluster & 100 ks & \S6.1 \\
30 Dor C & Superbubble & 300 ks &  \S6.2 \\
SS433/W50 lobes & Microquasar jets & 300 ks & \S7 \\ 
Sgr A* & Supermassive BH & 500 ks & GC paper (\S4) \\  
\hline
\label{tab:obs}
\end{tabular}
\end{center}
Note: 3.3 Ms total exposure. Some of the sources are described in our companion SNR/PWN (Reynolds et al. 2023) and GC papers (Mori et al. 2023).  
}
\end{table*}

\section{Galactic PeVatrons}

\begin{figure}[ht!]
    \centering
    \includegraphics[width=1.0\linewidth]{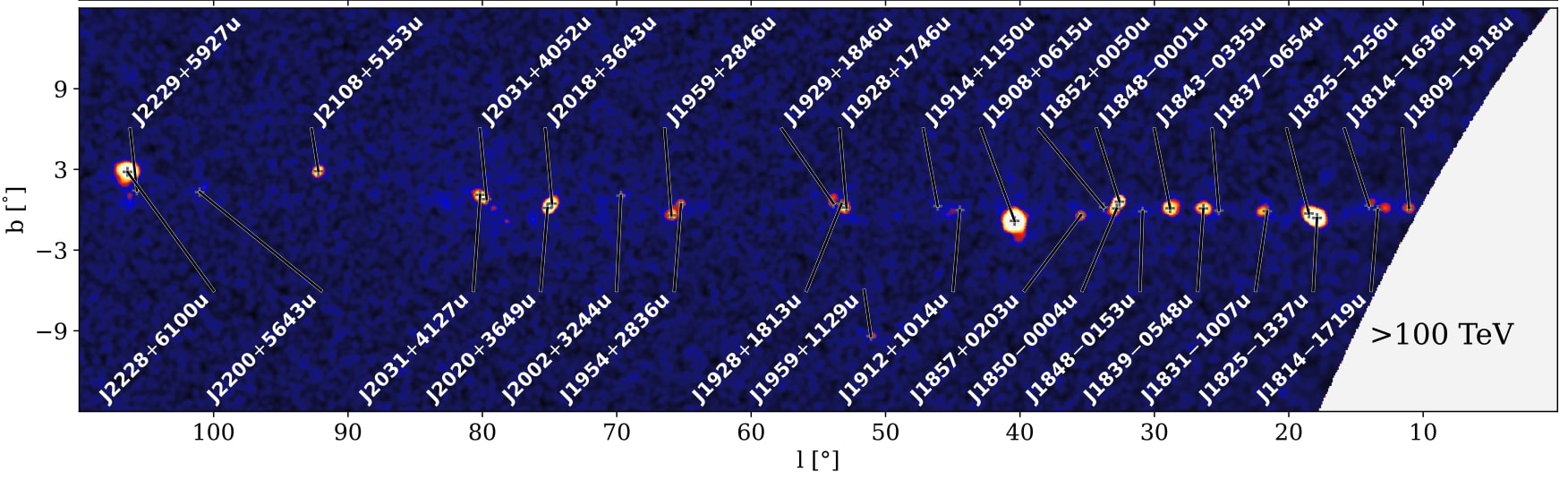}
    \caption{LHAASO $E > 100$ TeV significance map of the Galactic Plane ($10^\circ < l < 115^\circ$) excerpted from \citet{lhasso23}. 
    }
    \label{fig:lhaaso_map}
\end{figure}

With the advent of extensive air-shower arrays (EASAs) such as HAWC and LHAASO operating in the ultra-high energy (UHE) band ($E_\gamma\simgt100$ TeV), we are reaching the unprecedented regime of astroparticle physics by discovering PeVatrons that can accelerate CRs to PeV energies, far beyond the maximum energies reachable by the terrestrial particle accelerators ($\sim$ TeV). 
The presence of Galactic PeVatrons is a new, exciting reality since LHAASO detected 43 UHE sources in the Galactic disk \citep[Figure \ref{fig:lhaaso_map}; ][]{Cao2021, Cao2021b, Aharonian2021, lhasso23}. Similarly, HAWC and Tibet AS $\gamma$ detected  Galactic TeV sources above $\sim50$ TeV, overlapping with most of the LHAASO sources  \citep{Abeysekara2020, Amenomori2021}. 
Remarkably, some of the LHAASO sources were detected above 1 PeV, representing the most extreme known particle accelerators in our Galaxy. 
More Galactic PeVatrons are expected to be discovered as LHAASO and HAWC observatories accumulate more data over the next decade. PeVatrons in the southern sky remain unexplored where H.E.S.S reported several PeVatron candidates, such as HESS J1702$-$420A, whose gamma-ray spectra extend up to $\sim100$ TeV \citep{Abdalla2021}. The South Wide-field Gamma-ray Observatory (SWGO) is expected to discover PeVatrons in the southern sky, complementing the northern sky coverage of LHAASO and HAWC \citep{SWGO2022}. 
The recent discoveries of the UHE sources by the EASAs marked a paradigm shift in high energy astrophysics from ``Do PeVatrons exist in our Galaxy?" to ``What are the Galactic PeVatrons detected by EASAs?" and ``What is the contribution of Galactic PeVatrons to the UHE cosmic-ray population  
from the knee ($\sim 10^{15}$ eV) to the ankle ($\sim10^{18}$ eV)?".   


TeV gamma-ray sources, including the Galactic PeVatrons, are usually classified as  primarily leptonic or hadronic accelerators. 
A substantial fraction of energetic $e^{+/-}$ at GeV--PeV energies could be generated by PWNe and pulsar halos \citep{Cholis2022, LC2022}.  
A PWN is generally assumed to be a pure leptonic accelerator, with synchrotron and ICS emission from the expanding bubble of shocked, highly relativistic $e^{+/-}$ pulsar wind continuously injected by the pulsar. However, the possible presence of hadrons in the pulsar wind \citep[see, e.g.,][]{atoyan96,amato03,guepin20} has not been ruled out by present observations \citep{amato21}. 
In the hadronic accelerators, energetic CR ions diffuse out from the accelerator site, collide with any ambient dense material, and produce copious pionic showers which decay into neutrinos, $\gamma$-rays and (secondary) $e^{+/-}$. SNRs near clouds, massive star clusters, and superbubbles are considered the primary hadronic accelerators \citep{Cristofari2021, Aharonian2019}. Black holes in different mass scales have been recognized as another class of Galactic particle accelerators or possibly PeVatron candidates,  associated with X-ray binaries \citep{Kantzas2022}, microquasars \citep{Safi-harb2022}, and the supermassive black hole at Sgr A* \citep{2016Natur.531..476H}.  

\subsection{Multi-wavelength observations of Galactic PeVatrons with HEX-P and CTA}

While the EASAs such as LHAASO serve as PeVatron search engines, their UHE sources (above 0.1 PeV) are poorly localized or spatially extended due to their limited angular resolutions. The position and extent of the LHAASO sources usually have an uncertainty of $\sim0.1^\circ$ and $\sim0.3^\circ$, respectively \citep{Cao2021, Cao2023}.  
Therefore, observations at lower energy $\gamma$-ray band ($E_\gamma\simlt50$ TeV) by IACTs such as VERITAS, H.E.S.S, and MAGIC are crucial for resolving the UHE sources with $\simlt 0.1^\circ$ angular resolutions. However, the current IACTs have resolved only two UHE sources (LHAASO J2108+3651 and J1825-1326) into multiple distinct TeV sources so far, possibly because the UHE sources are largely extended, obscured by diffuse TeV emission in the Galactic Plane or fainter than the current IACT's sensitivity limits \citep{Aliu2014, Hess2020b}. Similarly, in the GeV band, \fermi-LAT sources were detected near some UHE sources, but the unique identification of GeV counterparts is often difficult due to the low angular resolution of \fermi-LAT. 
Resolving the UHE sources will need to wait until CTAO, the next generation IACT, becomes fully operational in the mid-2020s.  Observing Galactic PeVatrons is a key science project (KSP) for the CTAO mission, and it is recognized as one of the important science goals in the gamma-ray and astroparticle physics communities \citep{CTA2019}. CTAO is expected to resolve and localize the most energetic particle acceleration sites in the Galactic PeVatrons. 
 
\begin{figure}[h!]
    \centering
   \includegraphics[width=0.8\linewidth]{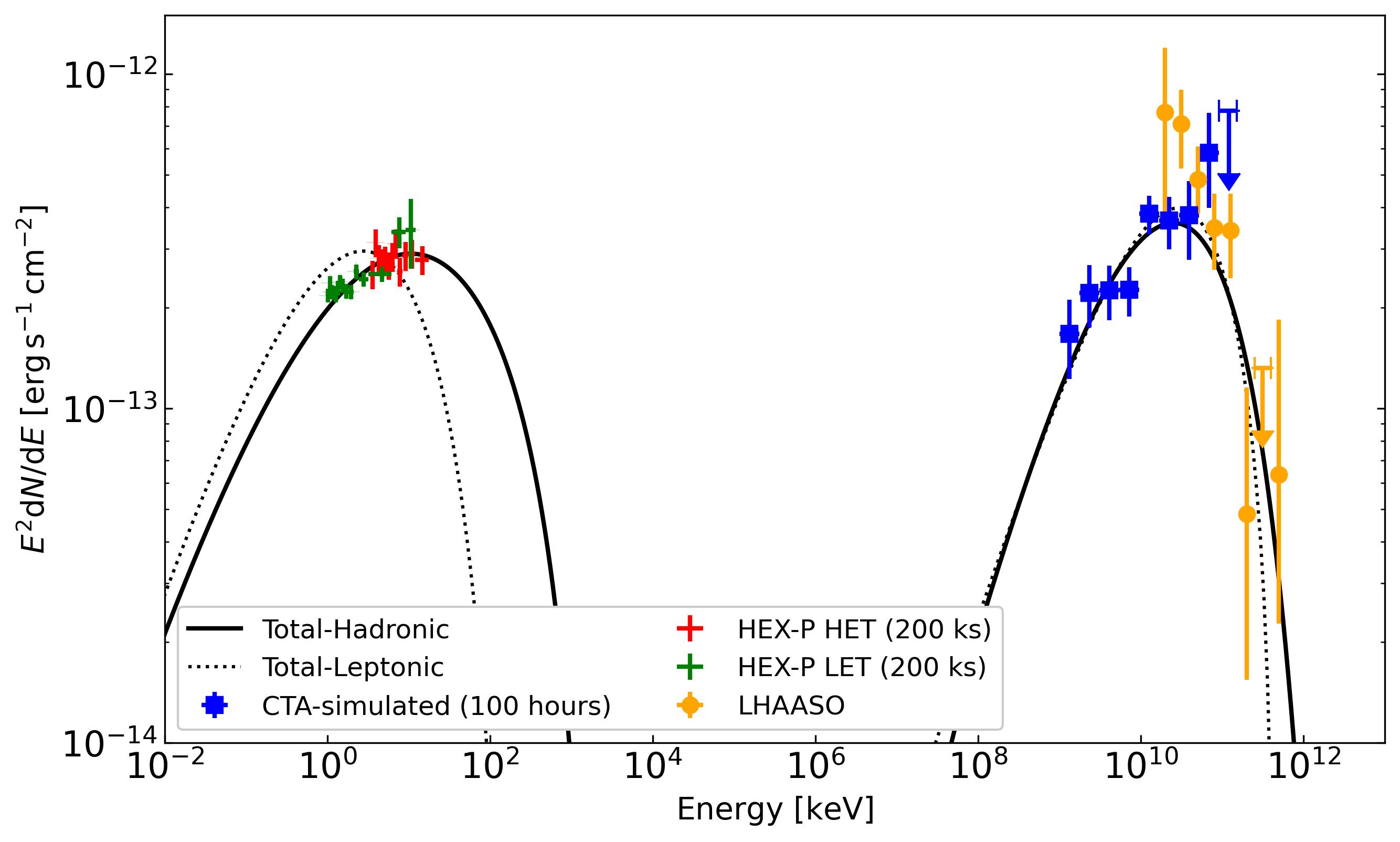}
    \caption{Multi-wavelength SED models for LHAASO J2108+5157 including both hadronic (black solid line) and leptonic (black dotted line) models. The flux points from LHAASO-KM2A \cite{lhasso23} are overlaid as yellow circles. The CTAO flux points (blue square) are from a 100-hour simulation using the Small-Sized Telescopes (SST) and Medium-Sized Telescopes (MST). The HEX-P LET and HET flux points, the green and red crosses, respectively, are from 200-ks simulations. 
    }
    \label{fig:J2108_SED}
\end{figure}

\begin{figure}[h!]
    \centering
   \includegraphics[width=1.0\linewidth]{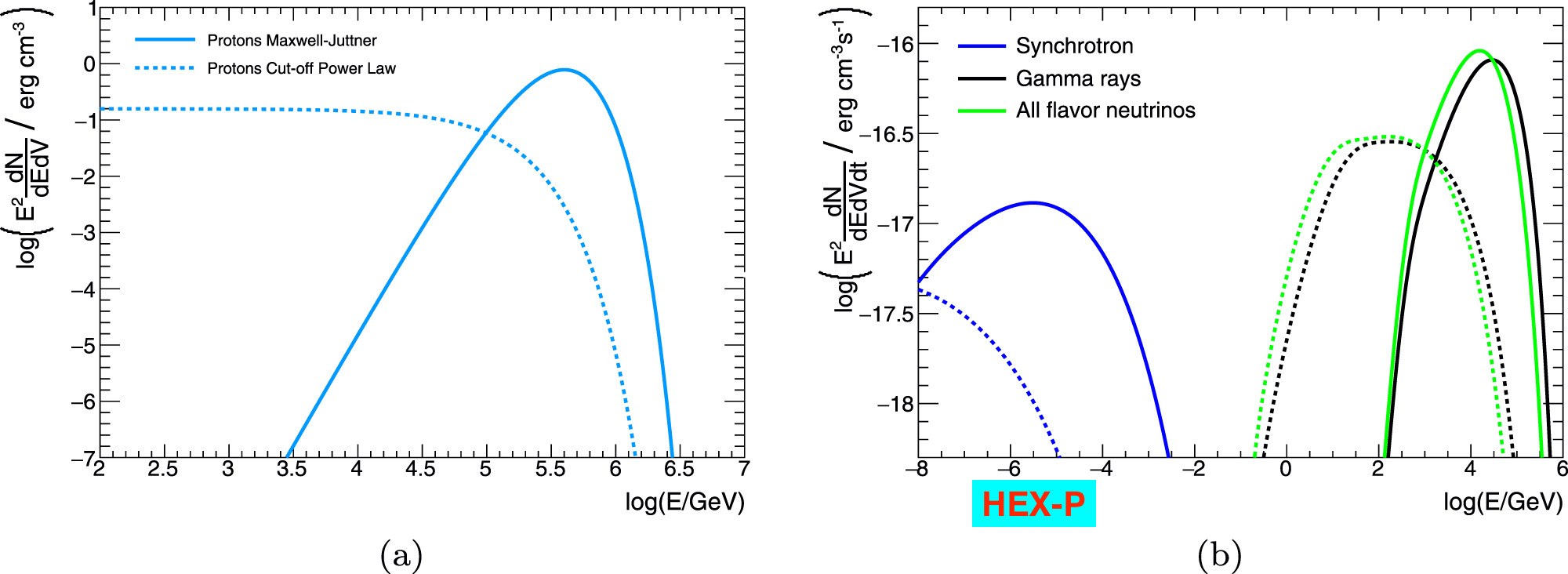}
    \caption{Primary proton energy distributions (left) and multi-wavelength SED models (right) for hadronic PeVatrons, excerpted from \citet{Celli2020}. Two types of primary proton spectral models are considered: Maxwellian-like  (solid lines) and power-law with an exponential cutoff (dashed lines). The synchrotron X-ray spectra are quite distinct between the two scenarios (assuming $B = 1$ mG). HEX-P will thus play a complementary role in constraining the primary CR spectra to the gamma-ray and neutrino telescopes. 
    }
    \label{fig:hadronic_SED}
\end{figure}

In the X-ray band, given its broadband coverage and high sensitivity up to 80 keV, HEX-P will be uniquely suited for exploring the nature and acceleration mechanisms of Galactic PeVatrons in synergy with CTA.  HEX-P will be optimal for detecting diffuse non-thermal X-ray emission and characterizing the known counterparts (e.g., PWNe, SNRs, star clusters) of the UHE targets. 
Both HEX-P and CTAO, reaching the regime of the highest energy  particles in the PeVatrons, are alike for their broad-band energy coverages and versatile functionality equipped with excellent angular, energy, and timing resolutions. CTAO will be able to resolve the LHAASO/HAWC sources with $\sim1'$ angular resolution and guide HEX-P in pinpointing their central engines and primary high-energy emission sites. Hard X-ray emission should originate from more compact regions (which will likely be covered by a single or few pointings with HEX-P) as observed from many of the known Galactic TeV sources (e.g., \citet{Coerver2019}). In both leptonic or hadronic particle accelerators, synchrotron X-ray radiation is expected from Galactic TeV sources through primary electrons and secondary electrons from pionic showers. Accurate measurements of the B-field are crucial since synchrotron radiation is typically the dominant particle cooling process for TeV-PeV electrons. 
Unfortunately, \nustar\ observations of many extended X-ray sources have been limited by high background contamination, often resulting in detections up to only 20 keV. Expanding on the previous X-ray + TeV observations (e.g., \nustar\ + HAWC + VERITAS, H.E.S.S + \suzaku\ surveys), multi-wavelength SED data obtained by HEX-P and CTAO will enable identifying the nature of Galactic PeVatrons in the 2030s.  An example, shown in Figure \ref{fig:J2108_SED}, depicts the fit of LHAASO and simulated CTAO SED data of one of the dark PeVatron accelerators (LHAASO J2108+5157) using both leptonic and hadronic models. The SED plot highlights a stark contrast in the X-ray fluxes and slopes within the HEX-P band. Moreover, Figure \ref{fig:hadronic_SED} displays multi-wavelength SEDs of a hadronic PeVatron with two different primary proton spectra. The synchrotron X-ray emission from secondary electrons provides the most sensitive diagnostics for determining the proton energy distribution, particularly in the cutoff 
 PeV energy band, unlike gamma-ray and neutrino SEDs  \citep{Celli2020}. 


In summary, HEX-P will play a crucial role in the multi-messenger investigations of PeVatron astrophysics. As part of the primary science program, HEX-P aims to observe five Galactic PeVatron candidates, while CTAO is expected to survey most of the PeVatron candidates through their KSP program. Very recently and excitingly, IceCube detected TeV--PeV neutrinos in the Galactic Plane as the first evidence of Galactic hadronic PeVatrons \citep{icecube2023}. HEX-P will survey various types of PeVatron candidates, including leptonic, hadronic, and dark accelerators (with no apparent low-energy counterparts or association with known astrophysical sources), paving the way  for more extensive PeVatron observations through the PI-led GTO or GO programs.

\section{Star clusters and superbubbles} 

Star clusters, observed in broad ranges of masses, ages, and stellar densities \citep{Pfalzner2009,Krumholz2019} might be found in different types of galaxies with high star-forming rates \citep{Whitmore2000,Adamo2020}. 
The evolution of the galactic environment is closely related to star formation, which is a key phenomenon that binds together important constituents from molecular gas to magnetic fields and cosmic rays in a close relationship. 
Star-forming regions both in the Milky Way and in starburst galaxies are known sources of broadband non-thermal radiation from radio to gamma rays, indicating the presence of relativistic particles. A significant part of massive stars is believed to evolve as clusters, which are gravitationally bound groups of stars of a common origin. Of particular interest are massive star clusters (MSCs) found in many regions of star formation. MSCs are sources of both thermal and non-thermal X-ray radiation, and they are considered effective cosmic ray accelerators \citep[e.g.,][]{Bykov2014,Aharonian2019}. The Milky Way contains a number of well-studied MSCs, such as NGC~3603 \citep{Drissen1995}, Westerlund~1/2 \citep{Clark2005,Zeidler2015}, Arches \citep{Figer2002}, and Quintuplet \citep{Figer1999}, which contain dozens and even hundreds of bright OB, Wolf-Rayet (WR), cool super- and hypergiant stars in the cluster cores of a parsec scale size.  

In recent years, MSCs and superbubbles have been recognized as one of the primary classes of hadronic accelerators, possibly accounting for some of the PeVatrons in our Galaxy \citep{Aharonian2019}. 
In hadronic accelerators, energetic CRs diffuse out  from the accelerator site (star cluster), collide with ambient medium and molecular clouds, and produce copious pionic showers which decay into neutrinos, gamma-rays, and electrons/positrons. 
MSCs contain a number of massive stars ($M > 20 M_\odot$) and sometimes form large-scale H II regions, the so-called superbubbles, by ionizing and heating the surrounding gas. 
Colliding winds in massive binaries (composed of OB and WR stars) can also accelerate particles efficiently and emit non-thermal high-energy radiation \citep{Pittard2020, Morlino2021}.  
An important clue to support the MSC origin of PeVatrons has arisen from a  recent discovery of diffuse gamma-ray sources around a handful of the MSCs and superbubbles. 
Remarkably, \hess\ and \fermi-LAT discovered the so-called $\gamma$-ray cocoons extending over $\sim$ 50--300 pc around two MSCs (Westerlund 1 \& 2) and two superbubbles (Cygnus and 30 Dor C in the LMC). A leptonic origin for the $\gamma$-ray cocoons is ruled out since TeV--PeV electrons will cool down before traveling over the cocoon size distance. 
There are two key features observed from the $\gamma$-ray  cocoons that suggest they are likely hadronic PeVatrons. First, \hess\ and \fermi-LAT detected hard $\gamma$-ray spectra with $\Gamma \approx 2$ up to $E\simgt10$ TeV with no spectral cutoff -- this is a potential signature of PeVatrons \citep{Yang2019}.  
Secondly, the CR proton densities decrease as $1/r$ where $r$ is the distance from the star cluster \citep{Yang2019}. The $1/r$ profile indicates that the CRs must be injected from the source  continuously over $\sim10^6$ years. SNRs alone are unlikely to produce $\gamma$-ray cocoons since an unreasonably high SNR birth rate of more than one per 100 years is required in the region \citep{Yang2019}. Similarly, the MSCs such as Arches and Quintuplet clusters have been proposed as alternative particle accelerators which cause diffuse TeV emission in the central molecular zone (CMZ) of the Galactic Center, in addition to the (currently dormant) supermassive black hole at Sgr A* \citep{Aharonian2019} 
Hence, the most plausible hypothesis is that MSCs continuously injected  TeV--PeV CRs into the ambient molecular clouds, along with episodic supernova explosions, over the past $\sim10^6$ years and formed the $\gamma$-ray cocoons and superbubbles extending over $\sim$50-300 pc \citep{Vieu2022, Gabici2023}. This hypothesis needs to be tested by multi-wavelength observations.


\subsection{Star clusters} 

In order to explore the formation of gamma-ray cocoons, firstly, it is essential to study how their central engines (i.e. MSCs) accelerate particles using X-ray and TeV data. Given that the gamma-ray cocoon profiles suggested continuous particle injections, their star clusters should be presently  accelerating particles. The TeV emission mechanism from MSCs is still unsettled partially due to the lack of associated non-thermal X-ray detection. TeV gamma-rays could be produced by hadronic interactions in the colliding winds or ambient molecular clouds or ICS emission due to the high radiation densities within the clusters. It remains uncertain whether the prominent MSCs such as Westerlund 1 exhibit non-thermal X-ray emission caused by colliding wind shocks, apart from the thermal X-rays from numerous massive stars. 
Note that \nustar\  data of Westerlund 1 and Cygnus OB associations are severely contaminated by background photons from nearby magnetar and X-ray binaries  \citep{Borghese2019, Mossoux2020}. From Westerlund 2, a hard X-ray component has been detected up to $\sim20$ keV by SRG/ART-XC but its origin remains undetermined between non-thermal ($\Gamma \sim 2$) and thermal plasma ($kT \sim 5$ keV) emission \citep{Bykov2023}. For the Arches cluster, variable non-thermal X-ray emission was detected by \nustar\ from its nearby molecular clouds, but not from the star cluster itself \citep{Krivonos2014}. 
So far, \nustar\ has identified non-thermal X-ray emission only from Eta Carina \citep{Hamaguchi2018}. Eta Carina has been considered an exception as it is the supermassive and most luminous binary star system in our Galaxy.

\begin{figure}[ht!]
    \centering
\includegraphics[width=1.0\linewidth]{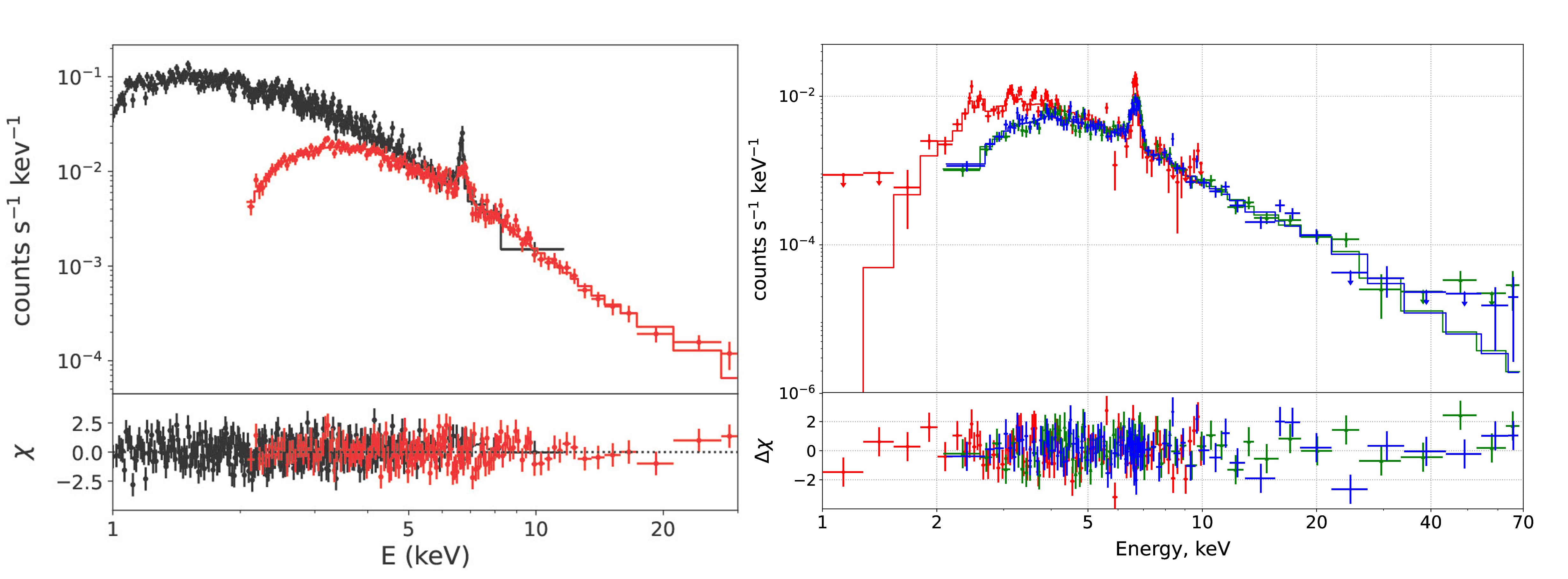}
    \caption{Left: Simulated \hexp\ LET + HET spectra of Westerlund 2 ($r < 90''$ around the core of the cluster). We input an absorbed APEC + power-law model, which was also fit to the simulated spectra. The spectral parameters used in the model ($kT = 2.6$ keV and $\Gamma = 2$) were determined by \nustar\ analysis of the same region. Note that the photon index is not well constrained by \nustar\ data due to the poor signal-to-noise ratio above 10 keV, also making the detection of a non-thermal emission component ambiguous. For comparison, we fit an absorbed  double-APEC model to the simulated HEX-P spectra and found that the second thermal plasma component yielded an unreasonable and unconstrained plasma temperature of $kT > 27$ keV. \\ 
    Right: Simulated spectrum of the Arches cluster emission expected from the circular region with $R=50''$ \cite{Krivonos2017, Kuznetsova2019}. The spectrum is presented for \hexp\ (red for LET and blue and green for HET) in the 1--70~keV energy band with the 150~ks exposure time. The non-thermal emission was simulated for a case of $\Gamma=1$ and the non-thermal flux of 7\% of the total flux in the 2--10~keV energy band. }
    \label{fig:msc_spec}
\end{figure}

Without detecting non-thermal X-ray emission, it remains elusive whether numerous massive stars and binaries can emit strong winds and accelerate particles collectively in the MSCs. Given its higher angular resolution and sensitivity than \nustar, HEX-P is best suited for resolving non-thermal X-rays from predominant thermal X-ray emission spatially and spectrally. HET is particularly important since the MSCs are usually crowded with other non-thermal X-ray sources such as magnetars and PWNe. 
A good example is Westerlund 1 where \nustar\ data are severely contaminated by X-ray photons from the magnetar CXOU J164710.2$-$455216  in the region. 
Overall, broad-band LET + HET spectra will be able to characterize both thermal and non-thermal X-ray components from star clusters with significantly reduced background levels than \nustar. For example, Figure \ref{fig:msc_spec} (left panel) displays simulated HEX-P spectra for Westerlund 2, whose non-thermal X-ray component is detected up to $\sim30$ keV. 
In addition to Westerlund 1\&2 and the Cygnus regions, HEX-P's GC survey program will cover Arches and Quintuplet star clusters which have been considered as one of the primary particle accelerators in the GC region \citep{Aharonian2019}. HEX-P observations of Orion and Carina nebulae will allow us to explore X-ray source populations and potential particle acceleration sites in younger star clusters  (Mori et al. 2023).

The star clusters Arches and Quintuplet, located in the GC, are known X-ray emitters: thermal and possibly non-thermal X-ray emission was detected by \chandra\ \citep{Law2004,Wang2006}. The nature of the non-thermal X-ray emission of the Arches cluster is not completely known. \cite{Tatischeff2012} mapped the molecular cloud in 6.4~pkeV Fe fluorescent line and made an assumption about collisional ionization by low energy CR (LECR) particles. Furthermore, \cite{Krivonos2014} studied the extended X-ray emission of the Arches complex, containing star cluster and nearby molecular cloud, at energies above 10~keV with \nustar\ data \citep[see also][]{Kuznetsova2019}. They showed that non-thermal emission is consistent with the X-ray reflection scenario, but also in broad agreement with the bombardment of the neutral matter by LECR protons. \cite{Clavel2014} showed that the X-ray flux from the Arches molecular cloud in the 6.4 keV line and continuum has been decreasing since 2012, which indicates cloud ionization from a possible Sgr~A* flare. Since then, studies of non-thermal emission from the Arches complex have not been carried out, and after a few years, a significant decrease in the flux can be expected. This opens a possibility to observe the Arches star cluster isolated, i.e. without strong contribution from non-thermal emission of the molecular cloud, and to detect possible non-thermal emission of the star cluster itself. 


To investigate the prospects of {\hexp} to discover the intrinsic non-thermal emission from the Arches star cluster, we consider the observed total X-ray emission in 2015--2016 with \nustar\ and {\it XMM-Newton} \citep{Krivonos2014,Kuznetsova2019}. The properties of the cluster's thermal emission are well known; for example, \cite{Kuznetsova2019} uses collisionally ionized plasma with a temperature of 1.95~keV and an unabsorbed 2--10~keV flux  $F_{apec}=1.16\times10^{-12}$~ergs~cm$^{-2}$~s$^{-1}$. Due to the unknown spectral form of the expected Arches cluster non-thermal emission, we consider a power-law model with a photon index  $\Gamma=1$ and $\Gamma=2$. To constrain the normalization of the power-law model, we imposed the 10--40~keV flux not to exceed the observed total emission of  $F_{\text{10-40 keV}}=4\times10^{-13}$~ergs~cm$^{-2}$~s$^{-1}$ within $50''$ from the cluster's center \citep{Kuznetsova2019}. For an exposure of 150~ks, we estimate the non-thermal flux to be detected at the significance of $8\sigma$ for both spectral indexes, with the uncertainty of spectral index at the level of 30\% and 10\% for  $\Gamma=1$ and $\Gamma=2$, respectively (see Fig.~\ref{fig:msc_spec} right panel for $\Gamma=1$). The simulated thermal flux of the Arches cluster $1.3\times10^{-14}$~ergs~cm$^{-2}$~s$^{-1}$ in 10--40~keV band can be considered as a threshold, above which the non-thermal component can be revealed.



\begin{figure}[h!]
\begin{center}
\includegraphics[width=0.8\textwidth]{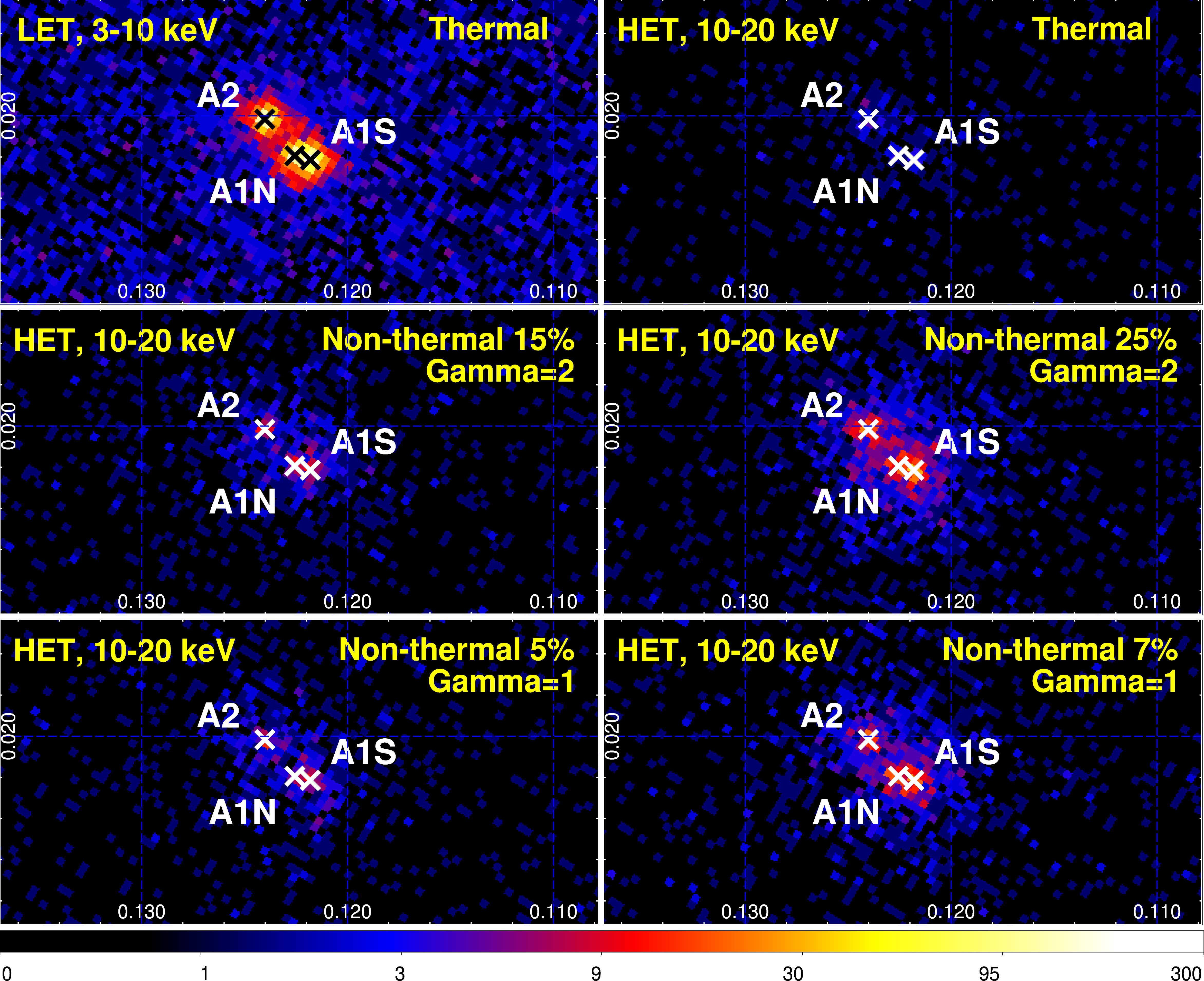}
\end{center}
\caption{150-ks simulations of various scenarios of thermal and non-thermal emission from the Arches cluster for LET and HET. The crosses correspond to the positions of the brightest point sources within the Arches cluster detected by \chandra\ \citep{Wang2006}. The assumed fraction of the non-thermal flux to the total flux in the 2--10 keV energy band and the photon index of the power-law component are labeled at the upper right corner of each figure. }\label{fig:arches}
\end{figure}

The angular resolution of the {\it LET} module will allow resolving bright point sources in the dense cluster's core. To model the spatial morphology of the cluster seen by {\it LET}, we utilize the spatial and spectral information of the bright sources A1N, A1S, and A2 in the Arches cluster detected with {\chandra} by \cite{Wang2006}. Figure~\ref{fig:arches} demonstrates the 150-ks LET simulation which allows resolving A2 from A1N and A1S. We conclude that LET will provide an opportunity to investigate the spatial morphology of the Arches cluster thermal emission in detail. 
In addition, Fig.~\ref{fig:arches} clearly demonstrates different source morphologies seen by HET for different cases of non-thermal emission likely to be emitted by point sources of the Arches cluster.


\subsection{Superbubbles} 

\begin{figure}[ht!]
    \centering    \includegraphics[width=1.0\linewidth]{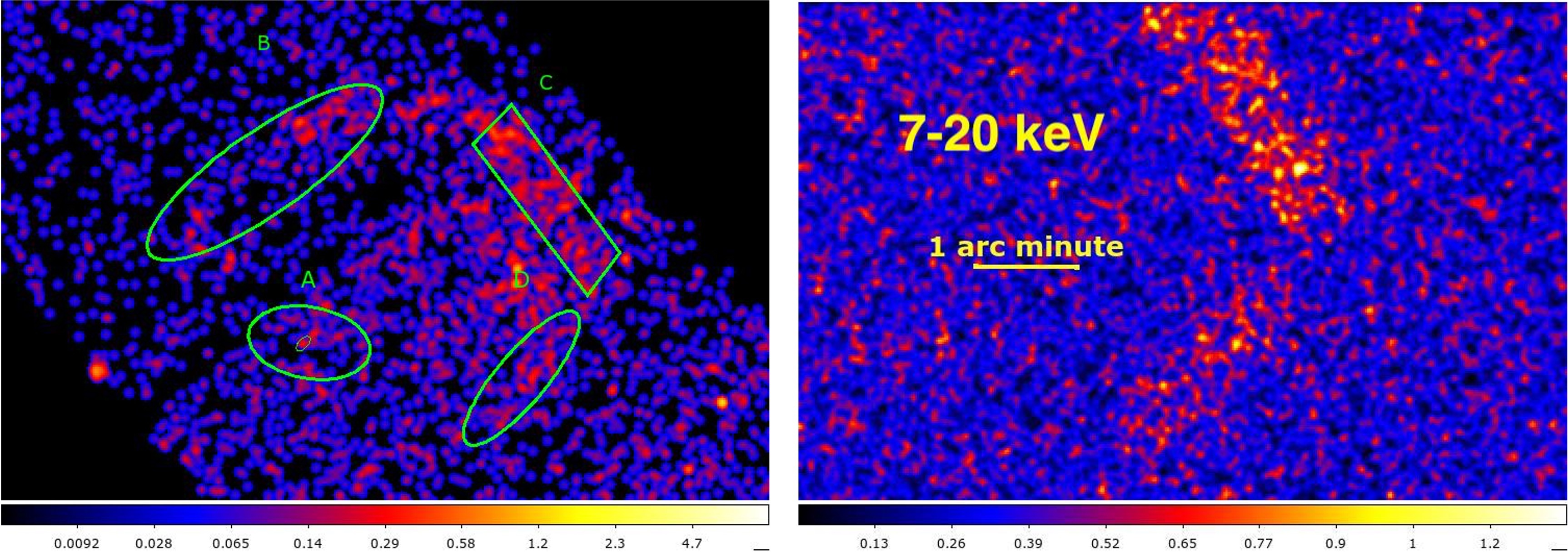}
    \caption{ Left: \chandra\ image around 30 Dor C in the 2 -- 7 keV band used as an input for SIXTE simulations. We input the best-fit spectral parameters from four shell-like regions, which correspond to the regions A--D shown in Figure 1 in \cite{bamba04}. 
    Right: Simulated HEX-P HET image of 30~Dor~C in the 7--20~keV band in a linear scale. One can see clear emission from its shells up to 20~keV with only 300~ks exposure. 
    }
    \label{fig:30DorC}
\end{figure}

While star clusters are the primary sites of accelerating and injecting particles, superbubbles extending over hundreds of parsecs emerge as a result of energetic particles propagating into the surrounding medium over millions of years. 30~Dor~C is the only superbubble detected with non-thermal X-rays \citep{bamba04} up to 20~keV \citep{lopez2020} and TeV gamma-rays \citep{hess15}. 
The non-thermal X-ray luminosity is 10 times brighter than that of SN~1006, and similar to that of RX~J1713.7$-$3946
\citep{nakamura12},
implying that 30~Dor~C hosts a powerful particle accelerator. 
\citet{bykov01,bykov06} proposed that
episodic supernova explosions in star clusters can generate multiple shock waves and their  interactions within the superbubbles result in  energetic particle acceleration. This idea is supported by the \xmm\ detection of a young SNR inside 30~Dor~C \citep{kavanagh15}. However, 
we still lack a global picture of which physical parameters control the particle acceleration mechanism since no other superbubbles have non-thermal X-ray emission detected yet \citep{yamaguchi10}. 
The acceleration mechanism of superbubbles seems to operate differently from that of SNRs since the superbubble is filled with less dense, optically-thin hot plasma. Due to the smaller Mach numbers of the SN shocks in the superbubbles, the shock waves expand without deceleration until they abruptly slow down upon colliding with the surrounding dense gas. 
Given the complex interactions between SN shock waves and ambient gas, it is challenging to model the maximum energy of accelerated particles and their energy evolution.  
To determine the contribution of superbubbles to the Galactic cosmic-ray populations, it is crucial to measure the maximum electron energies  through broadband non-thermal X-ray spectra from superbubbles. HEX-P observations of superbubbles, leading to measuring their X-ray spectral indices and roll-off energies, will allow us to elucidate how their particle acceleration processes are different from those of SNRs (e.g., Cas A and Tycho). For example, a simulated HEX-P HET image of 30 Dor C is presented in the right panel of Figure \ref{fig:30DorC}. Note that 30 Dor C will be covered by the 300 ks HEX-P observation of SN1987A in the same FOV. The simulation clearly shows  hard X-ray emission up to 20 keV with only 300 ks (compared to the \nustar\ observations with $\sim3$ Ms yielding similar results \citep{lopez2020}), highlighting HEX-P's capability of detecting extended sources in the hard X-ray band. 

\section{SS433 / W50 lobes}
\label{sec:ss433}

\begin{figure}[ht!]
    \centering
    \includegraphics[width=0.7\linewidth]{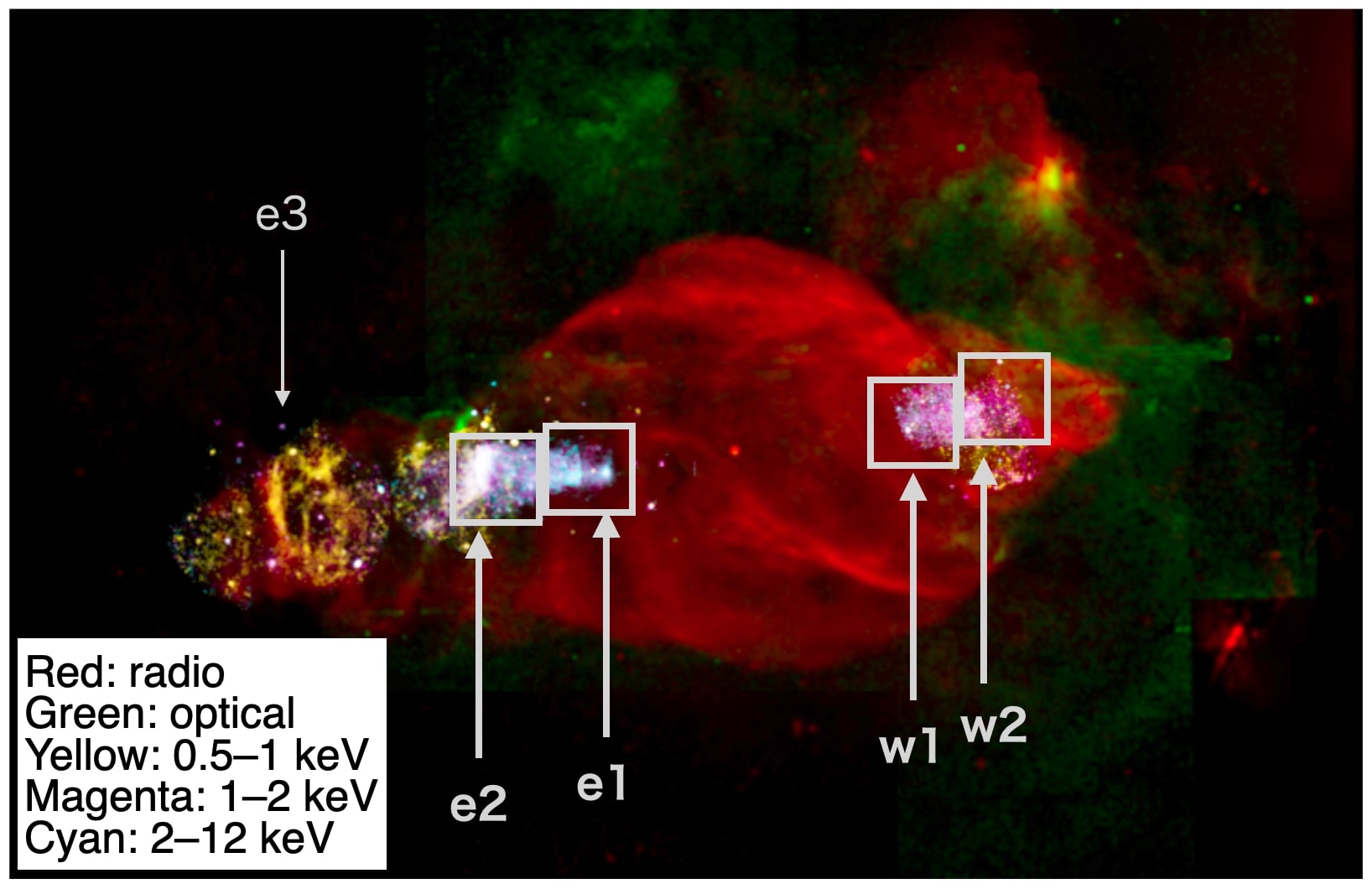}
    \caption{Entire image of the SS~433/W50 system, taken from \cite{safi-harbHardXrayEmission2022}. 
    White boxes show the FoV of HEX-P and our proposed pointings.
    }
    \label{fig:ss433_image}
\end{figure}

The Manatee Nebula W50 is one of the most prominent radio sources in the sky, associated with the microquasar SS 433  located at a distance of 5.5 kpc (Figure~\ref{fig:ss433_image} and, e.g., \citet{Dubner1998}). The bipolar jets launched from SS 433 interact with the ISM and W50 nebula, producing distinct features in multiwavelength bands.
Of particular interest are knot-like structures at both the eastern and western lobes, referred to as e1--2 and w1--2, recently shown to be sites of TeV gamma-ray emission \citep{Hawc2019,lhasso23}. 
The initial ASCA/ROSAT/RXTE/XMM surveys identified distinct X-ray knots in the eastern (e1--e3) and
western (w1--w2) lobes \citep{safi-harbROSATASCAObservations1997,safi-harbRossiXRayTiming1999,brinkmannXMMNewtonObservationsEastern2007} with non-thermal X-ray emission dominating the inner regions, likely
synchrotron radiation from accelerated electrons. 
Only e3 has an apparent counterpart in the radio band known as the eastern ``ear". 
The HAWC discovery of TeV gamma-ray emission from the eastern and western lobes, recently also detected by LHAASO \citep{lhasso23}, indicated that the SS 433/W50 system could represent another class of extreme particle accelerators powered by microquasar jets  \citep{Hawc2019}. 

The detection of TeV emission by HAWC has motivated extensive studies of the W50 system, including multi-wavelength surveys below the TeV band, theoretical modeling, and numerical simulations.
Among various follow-up observations, \xmm\ unveiled that there exists a relatively compact structure at the e1 knot, dubbed the ``head" region (Figure~\ref{fig:ss433_image}).
Broadband X-ray spectra with \xmm\ and \nustar\ 
were characterized by a power-law model with $\Gamma \sim 1.5$ up to $\sim$30 keV in the head region \citep{safi-harbHardXrayEmission2022}. 
In the western lobe, using the archival \chandra\ data  \citep{moldowanMultiwavelengthStudyWestern2005}, \citep{kayamaSpatiallyResolvedStudy2022} recently extracted 
a detailed profile of spectral parameters along the western lobe (w1--w2), revealing that non-thermal X-ray emission begins at w1 (where the particle acceleration is initiated and most energetic) and becomes gradually softer toward w2 due to synchrotron cooling. In the radio band, an extensive VLA survey in the 1.4 GHz band was conducted over the entire W50 system, mapping synchrotron radiation emitted by lower-energy (GeV) electrons \citep{sakemiEnergyEstimationHighenergy2021}. 
As shown in Figure~\ref{fig:ss433_image}), the W50 ``mini-AGN" system manifests all elements of astrophysical jets: acceleration sites (the inner lobes); particle propagation/cooling along the jet; and thermalization at the termination region. Motivated by the HAWC discovery in 2018, a handful of particle acceleration models have been developed, including leptonic and hadronic SED models \citep{Sudoh2020,kimuraDecipheringOriginGeV2020} as well as MHD simulations \citep{ohmuraContinuousJetsBackflow2021}.

Despite the extensive X-ray surveys with \xmm, \chandra, and \nustar, we are still not at the stage of fully testing these theoretical model predictions, let alone deciding the origin of TeV emission between the leptonic and hadronic acceleration mechanisms. The \nustar\ data of the head regions (e1 and w1) are severely contaminated by ghost-ray background photons from SS~433 \citep{safi-harbHardXrayEmission2022} above $\sim 30$ keV. The TeV emission needs to be resolved with $\simlt 20''$ angular resolution so that it can be compared well with the X-ray and radio data. Apparently, we need multi-zone, multi-wavelength observations and modeling tied together, in order to completely determine how particles get  accelerated and propagate, while cooling, throughout the bipolar jets. As mentioned repeatedly for other types of particle accelerators, HEX-P and CTAO will make the highest impact on understanding this complex system, including local MHD phenomena such as magnetic-field amplification and knot formation.

\subsection{Scientific objectives with HEX-P}

\begin{figure}[ht!]
    \centering
    \includegraphics[width=0.9\linewidth]{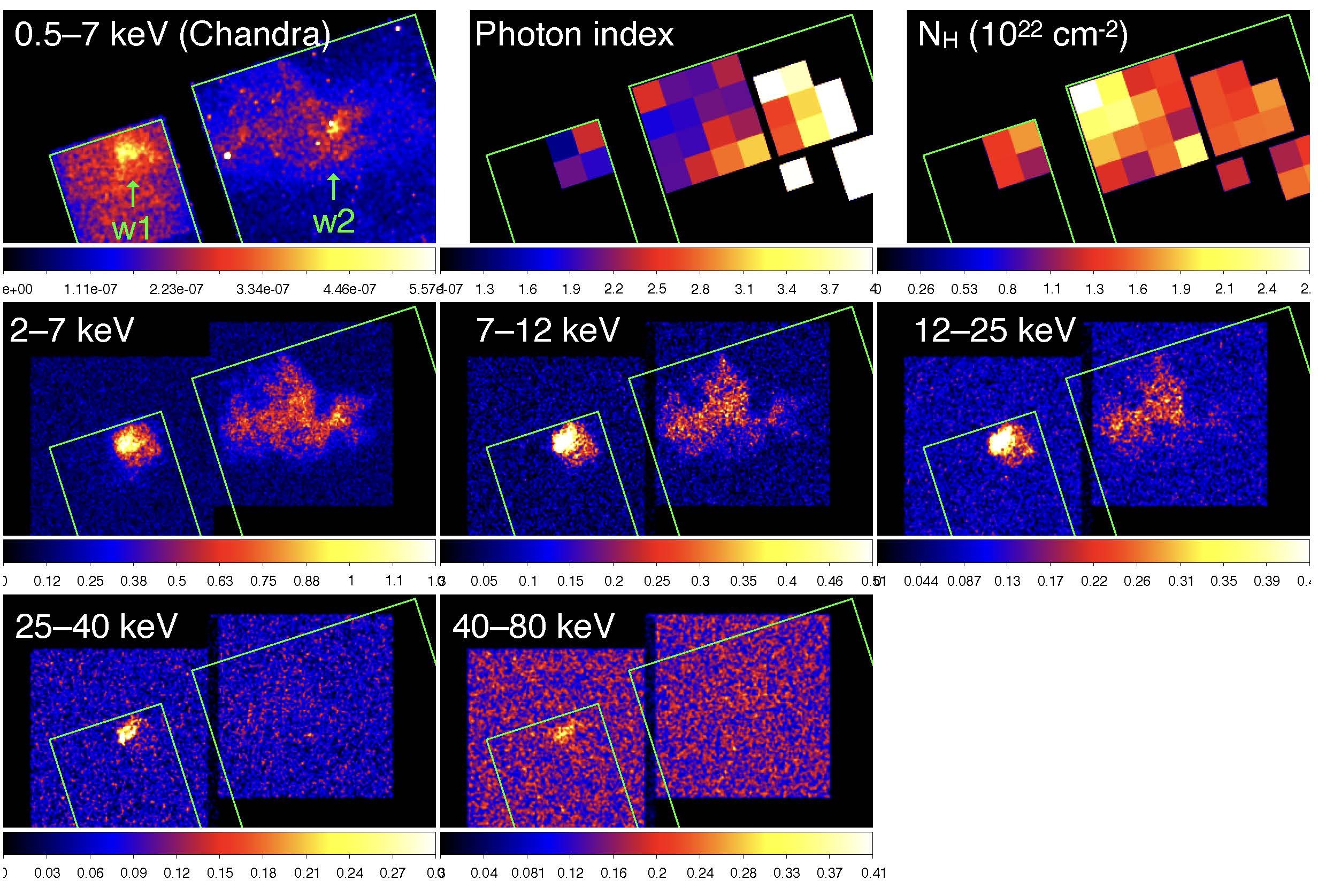}
    \caption{
    Top: Input \chandra\ flux, photon index, and column density maps of SS433/W50 western lobe in 0.5--7 keV.
    Middle and Bottom: Simulated HET images in five different energy bands of w1--2 with exposure of 75 ks. 
    Green lines show the FoV of \chandra\ ObsID 3843.
    }
    \label{fig:ss433_sim_image}
\end{figure}

A HEX-P X-ray survey of the SS~433/W50 system will be able to resolve the spectral and spatial profiles and will offer a unique opportunity to investigate particle acceleration by microquasar jets and their interactions with the surrounding environment. During the primary science program, HEX-P plans to survey the SS~433/W50 region with four observations pointing at the e1, e2, w1, and w2 knots, as shown in Figure \ref{fig:ss433_image}.  
To assess the feasibility of HEX-P observations, we conducted simulations with the SIXTE and XSPEC packages. For instance, Figure~\ref{fig:ss433_sim_image} illustrates simulated HET images of the western lobe in different energy bands with an exposure of 75 ks. The simulations use as input the \chandra\ flux image in 0.5--7 keV and spatial distributions of spectral parameters, $N_H$, flux, and $\Gamma$, which were adopted from Figure~6 in  \cite{kayamaSpatiallyResolvedStudy2022}. 
Note that the w1 knot, which exhibits a harder X-ray spectrum, remains clearly visible up to $\sim80$ keV, while the softer w2 knot can be detected up to $\sim25$ keV. Hence, HEX-P will have the capability of fully characterizing the X-ray spectral and spatial distributions, 
 tracking the evolution of relativistic particles along the jets. 

(1) Determining the particle acceleration mechanism:   
The origin of the TeV emission from the inner lobes remains uncertain since the current multi-wavelength SED data do not provide conclusive evidence for distinguishing between the leptonic and hadronic models \citep{Kimura2020}. Similar to other TeV sources, gamma-ray information alone is insufficient to 
 discriminate between the leptonic and hadronic scenarios or to constrain model parameters. Instead, broad-band X-ray spectroscopy and morphology studies with HEX-P, along with CTAO, will be critical in elucidating  the origin of the TeV emission and constraining key model parameters such as magnetic field ($B$) and electron/proton spectral indices. 
The morphology data obtained by HEX-P, capturing non-thermal X-rays, will enable us to spatially correlate synchrotron X-ray emission with molecular clouds   \citep{yamamotoPhysicalPropertiesMolecular2022} and radio features \citep{sakemiEnergyEstimationHighenergy2021}. Note that the molecular cloud  and radio maps track the target material distribution for hadronic interactions and the magnetic field distribution, respectively.  
The identification of hot spots in hard X-rays, coinciding spatially with molecular clouds and high-density optical filaments, would lend support to the hadronic case where synchrotron X-ray radiation originates from secondary electrons (produced by proton-proton collisions and subsequent pion decays) \citep{Kimura2020}.  
Conversely, in the leptonic case, a more gradual X-ray spectral softening is expected from the acceleration site (w1/e1) to the termination region (w2/e2 or e3) as propagating electrons cool down via synchrotron radiation \citep{Sudoh2020}.  
A combination of multi-wavelength SED analysis and the localization of energetic electrons, for which the hard X-ray band coverage of HEX-P plays a crucial role, can distinguish between the leptonic and hadronic scenarios robustly \citep{Kimura2020}.  

(2) Constraining acceleration efficiency along the jets: Measuring the cutoff energy ($E_c$) in the synchrotron X-ray spectrum offers direct insights into the particle acceleration mechanism by constraining 
an ``acceleration efficiency factor'' $\eta_{\rm acc} \equiv c\,\tau_{acc}/r_L$, where $\tau_{acc}$ is the acceleration time to energy $E$ and $r_L$ the Larmor radius \citep{Sudoh2020}.  That is,
$\eta_{\rm acc}$ is the dimensionless acceleration time, measured in
units of the inverse of the Larmor frequency.  
Particularly, in the case of DSA and cooling-limited electrons, the combination of $E_c$  and shock velocity provides a robust means to derive $\eta_{\rm acc}$ independent of the $B$-field. 
In the standard diffusion scenario, $\eta_{\rm acc}$ is characterized by the gyrofactor $\eta_g$, the ratio of the mean free path of a particle to its gyroradius ($\eta_g = 1$ is the ``Bohm limit"). Ultimately, $\eta_{\rm acc}$ and $E_c$ can be used to determine the maximum energies of accelerated particles, assuming an age-limited case. Of particular importance is the determination of whether $\eta_{\rm acc}$ is smaller than  $10^2$, as it implies the acceleration of particles to PeV energies in the formalism of \citep{Sudoh2020}. 
Previous observations with \nustar\ detected non-thermal X-ray emission in the head regions, where the particle acceleration is considered to be most active, up to $\sim$20--30 keV, but a cutoff was difficult to determine because of the high background level. In contrast, HEX-P holds a great  premise of extending its sensitivity up to 80 keV, enabling  the more accurate determination of cutoff energies and further constraining the acceleration efficiency factors. 

\begin{figure}[ht!]
    \centering
    \includegraphics[width=1.0\linewidth]{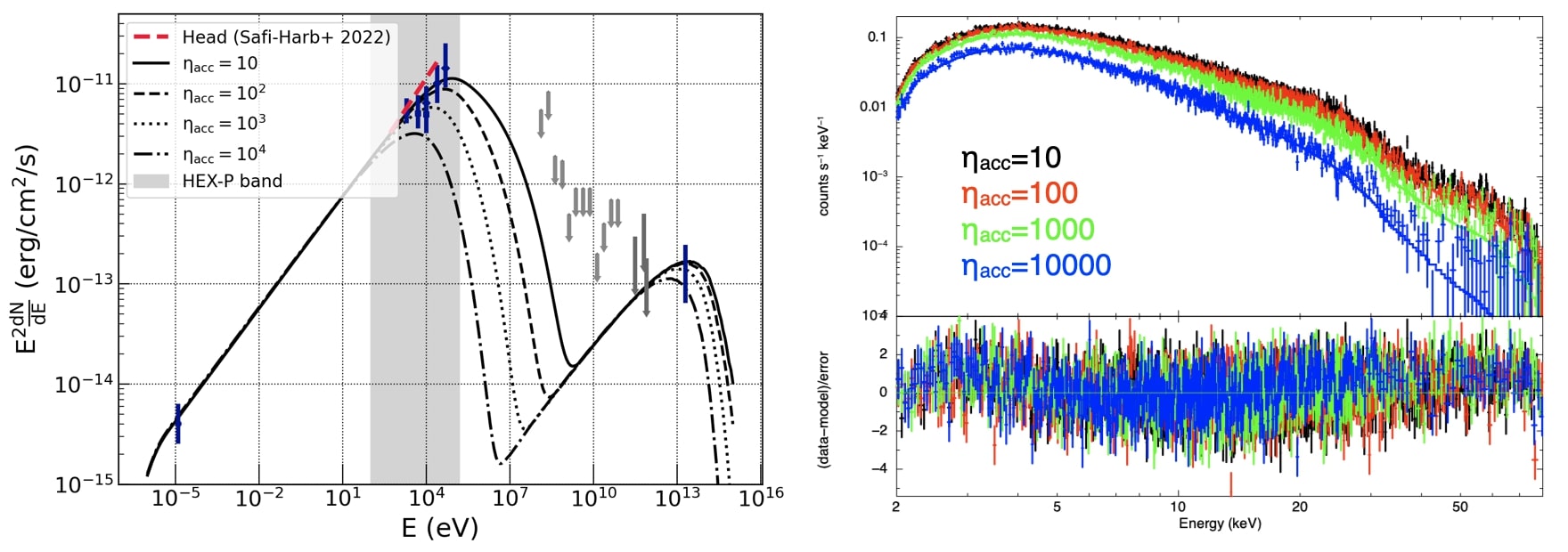}
    \caption{
    Left: multiwavelength SED of e1 and theoretical models with different acceleration efficiency ($\eta_{\rm acc}$) by \cite{Sudoh2020}. 
    Right: simulated HET spectra of e1 with exposure of 75 ks.
    }
    \label{fig:ss433_sed}
\end{figure}

In order to demonstrate the unique capabilities of HEX-P determining parameters related to particle acceleration, we 
 conducted simulations based on the leptonic model developed by \cite{Sudoh2020}, covering a wide range of $\eta_{\rm acc}$ values. For the case of DSA, the $\eta_{\rm acc}$ parameter and the gyrofactor $\eta_{\rm g}$ are related: $\eta_{\rm acc} \simeq 10^2 (\eta_g/2) (v_{\rm sh}/0.26c)^{-2}  $ (see \cite{Sudoh2020} for details).
Figure~\ref{fig:ss433_sed} displays the simulated HET spectra of the e1 knot with an exposure time of 75 ks.
We found that a spectral cutoff energy can be measured  with $< 10$\% accuracy, which is sufficient to distinguish between $\eta_{\rm acc} = 10$, $10^2$, $10^3$, and $10^4$. 
In the w1 knot where its X-ray flux is fainter, a 75-ks HEX-P observation can unambiguously determine whether the acceleration is in the Bohm ($\eta_g \sim 1$) or non-Bohm ($\eta_g \gg 1$) regime. A longer exposure time will constrain $\eta_{\rm acc}$ values in the w1 knot similarly to the e1 knot. 
The determination of acceleration efficiency within the microquasar jets will provide valuable insights among accelerators with different scales, such as $\eta_g \sim 1$ (the most efficient acceleration case) at SNR shells (e.g., \cite{tsujiSystematicStudyAcceleration2021a}) and $\eta_g \sim 10^6$ at AGN jets (e.g., \cite{araudoPARTICLEACCELERATIONMAGNETIC2015,inoueBARYONLOADINGEFFICIENCY2016}). 
By combining HEX-P and CTAO observations with the existing radio data, it will be possible to perform the most 
 refined multiwavelength SED analyses at various X-ray knot locations, as shown in Figure~\ref{fig:ss433_sed} (left panel), at sub-arcminute scales. This panchromatic approach will yield mapping of the distribution of acceleration efficiency and boost our understanding of particle acceleration and evolution along the microquasar jets/lobes.

(3) Knot formation in the microquasar jets/lobes: 
The formation and evolution of X-ray knots along the jets, not only in the case of W50 but also in AGN jets, remain uncertain and represent a long-standing question in astrophysics. 
Theoretical studies have proposed that the interaction of the jets with the ambient medium produces knot-like structures, 
as demonstrated by MHD simulations \citep{ohmuraContinuousJetsBackflow2021}. The proximity of W50 provides a unique opportunity for correlating the X-ray knots and known ambient features such as molecular clouds and filaments. 
Furthermore, it is not fully understood whether the knot sizes are determined by radiation loss, adiabatic cooling, magnetic field amplification, and re-acceleration \citep{Sudoh2020}. A recent X-ray study suggested synchrotron cooling, in combination with amplified B-fields, dominates in several locations of the western lobe (e.g., w2), in order to reproduce the X-ray spectral profile obtained by \chandra\ observations  \citep{kayamaSpatiallyResolvedStudy2022}. 
If synchrotron cooling is indeed predominant in most or all of the X-ray knots, HEX-P will contribute to the determination of local B-fields by measuring the presumably energy-dependent sizes of the X-ray knots. The broad-band X-ray morphology traces the
production site and cooling timescales of TeV-PeV electrons  by detecting synchrotron burn-off effects. 
By directly measuring local B-fields, HEX-P can uniquely investigate the origin of X-ray knot formation and test the leading theoretical hypothesis of B-field amplification \citep{Sudoh2020}. Consequently, a HEX-P survey of W50 will provide valuable insights into the processes involved in knot formation and contribute to our fundamental understanding of astrophysical jets. 




\section{TeV gamma-ray binaries} 

\begin{figure}[ht!]
    \centering
    \includegraphics[width=0.7\linewidth]{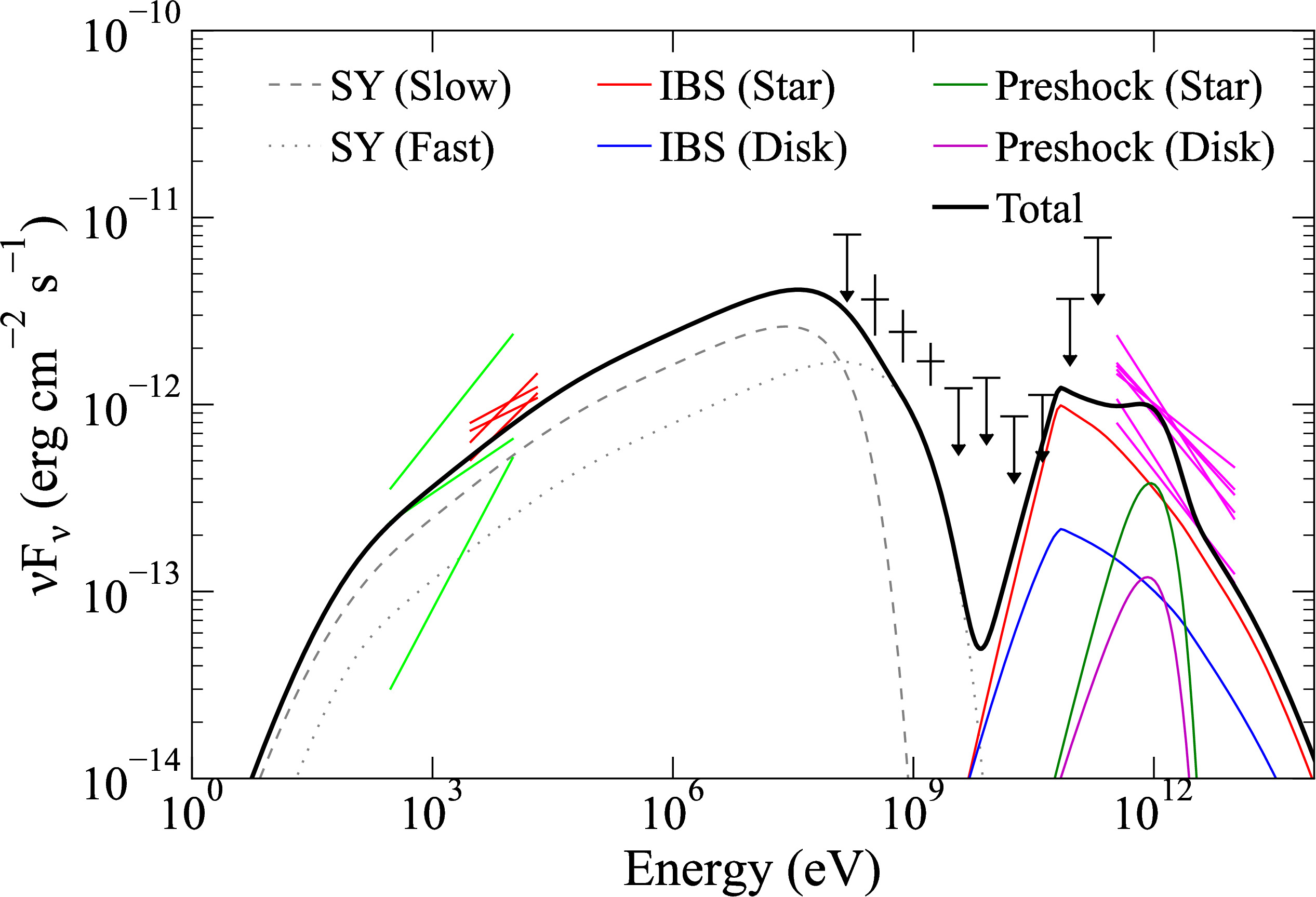} \\
    \caption{Broadband SED of the TGB HESS~J0632+057 and an IBS model \citep[figure taken from][]{Kim2022}. SEDs measured with \swift\ (lime), \nustar\ (red), and H.E.S.S. (magenta) are displayed with line segments with each line representing a measurement at an orbital phase. The data points at $\sim$GeV are LAT-measured fluxes, which may include the putative pulsar's magnetospheric emission. The dashed or colored curves show various model components and the thick curve is the sum of them.
        }
    \label{fig:J0632SED}
\end{figure}

TeV gamma-ray binaries (TGBs) are unique binary systems composed of a compact object and a massive companion, typically an O- or B-type star. To date, fewer than 10 TGBs have been discovered within our Galaxy, with one additional TGB in the LMC \citep[][]{Corbet2016}. In three of these TGBs, the compact object was identified as a pulsar. While TGBs belong to a subclass of high-mass X-ray binaries (HMXBs), they possess distinct properties, notably their predominantly non-thermal SEDs peaking above MeV energies. 
Except for the intense optical blackbody (BB) emission from their companions, multi-wavelength SEDs of the known TGBs exhibit a double-humped feature (e.g., Figure~\ref{fig:J0632SED}). The low-energy hump, observed from radio to X-ray band, arises from synchrotron radiation emitted by energetic electrons present in either the jets of a BH \citep[microquasar model, e.g.,][]{Marcote2015} or  the intra-binary shock (IBS) formed by interactions between the pulsar and companion star winds  \citep[pulsar-wind model, e.g.,][]{Dubus2013}. The high-energy bump, observed in the $\ge$100\,GeV band, is due to ICS between relativistic electrons and BB photons from the companion. The X-ray emission above $\sim10$ keV and the $>$TeV emission imply the presence of highly energetic (TeV) particles within TGBs. Furthermore, the broadband emission from TGBs exhibits strong orbital modulation, which can be attributed to the Doppler beaming of shocked particle flow and orbital variations of the ICS geometry. 

For the TGB systems in which the compact object is known to be a pulsar \citep[PSR~B1259$-$63, PSR~J2032+4127, and LS~I~+61$^\circ$~303;][]{Johnston1992,Abod2009Sci,Weng2022}, an IBS in the  pulsar-wind scenario are responsible for the broadband non-thermal emission. These systems all contain a Be-type companion with an equatorial disk. Broadband emission properties of these TGBs have been best studied for the archetypal object PSR~B1259$-$63. The pulsar crosses the disk at orbital phases near the periastron, and the pulsar-disk interaction produces dramatic X-ray and TeV flares at the crossings, accompanied by delayed  orphan GeV flares. The physical mechanisms responsible for these flares are not well understood yet but are speculated to be related to the pulsar--disk interaction which could enhance seed photon density and/or (partial) disruption of the IBS by the disk. \nustar\ observations of PSR~B1259$-$63 during such orbital intervals (including the GeV flare periods) revealed that the spectra were well fit by a hard $\Gamma_X=1.5$ power-law model, whereas the source spectra were softer ($\Gamma_X\approx$1.8--2.0) at the periastron and disk-crossing phases \citep[][]{Chernyakova2015}. These observed spectral variations are likely caused by a change in the particle injection spectrum and/or enhanced cooling at the disk-crossing phase. In the former case, we expect a power-law X-ray spectrum extending to $\sim$ MeV energies, where the synchrotron cooling break is expected in TGBs  \citep[e.g.,][]{An2017}. In the latter case, however, the amplified B-field of the IBS, caused by  compression from the circumstellar disk \citep{Tokayer2021}, may increase the synchrotron cooling rate, leading to a spectral break in the X-ray band. Thus, accurate measurements of the X-ray and gamma-ray spectral shapes are important to elucidate the particle acceleration and flow processes within relativistic shocks.

\subsection{Scientific objectives with HEX-P} 

\begin{figure}[ht!]
    \centering
    \includegraphics[width=1.0\linewidth]{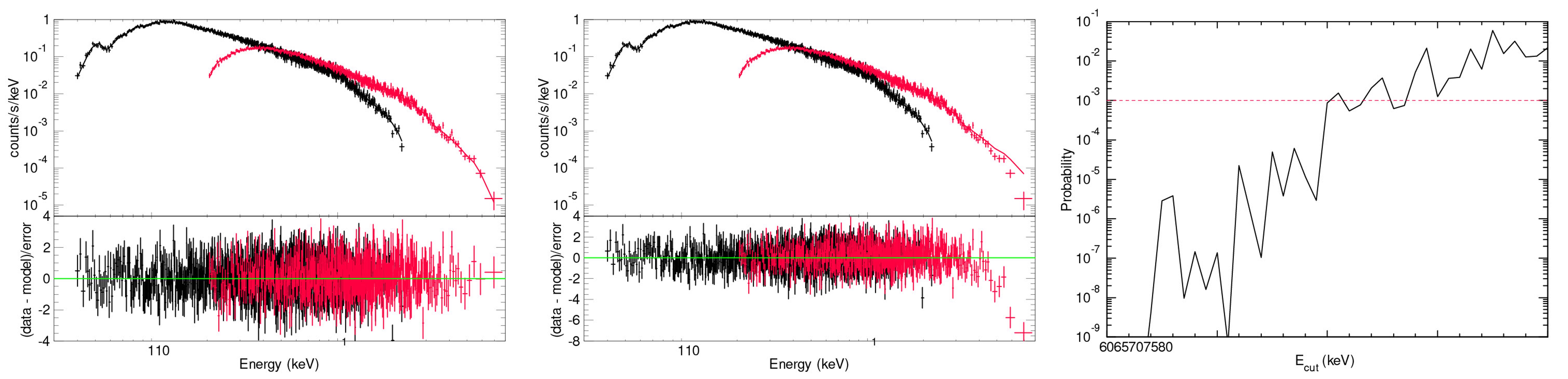}
    \caption{Results of simulations for 50-ks HEX-P observations of PSR~B1259$-$63 at a disk-crossing phase. Simulated HEX-P spectra with a cutoff at 60\,keV, along with the best-fit cut-off power-law and simple power-law models, are displayed in the left and middle panels. The residuals in the bottom show that the spectral cutoff at 60 keV is clearly detectable by HEX-P. $F$-test probabilities for detecting an exponential cutoff for a range of $E_{\rm cut}$ values are presented in the right panel where the red horizontal line marks $p=10^{-3}$. 
        }
    \label{fig:B1259disk}
\end{figure}

Previous \nustar\ observations in the soft state of PSR~B1259$-$63 were consistent with a simple power-law model to $79$\,keV. However, the sensitivity of \nustar\ data was limited in detecting a break or cutoff at $>$30\,keV due to the high background level. Given the large effective area, higher angular resolution, and reduced background, HEX-P will significantly improve the current measurements of the X-ray spectra during the disk crossing phases. Figure~\ref{fig:B1259disk} displays simulation results for a 50-ks HEX-P observation during a disk-crossing phase. For the simulations, we used an exponential-cutoff power law ($K (E/1\rm \ keV)^{-\Gamma_X}\mathrm{exp}(-[E/E_c]^{\alpha})$), where $K$ and $\Gamma_X$ were obtained from the \nustar\ results of PSR~B1259$-$63 at the crossing phase \citep[power law with $\Gamma_X=1.84$ and $F_{1-10\rm\ keV}=2.84\times 10^{-11}$\,\fluxcgs;][]{Chernyakova2015}. We held the exponential index $\alpha$ fixed at 5 and varied $E_c$. We then fit the simulated spectra with a power-law and an exponential-cutoff power-law model, and we employed the $F$-test to discern between the two models. Our results, shown in the right panel of Figure ~\ref{fig:B1259disk}, suggest that HEX-P will be capable of detecting an exponential cut-off at $E_c\le 70$\,keV with 50-ks exposure, and even a milder cut-off (e.g., a smaller exponential index) can be detectable with HEX-P. Moreover, the more accurate spectral measurements and identification of spectral features at different orbital phases can contribute to distinguishing between various emission components in TGBs, such as the IBS {\it vs} preshock emission \citep[e.g.,][]{Kim2022}.

Furthermore, HEX-P offers opportunities for other interesting studies. Since TGBs belong to a rare and intriguing class of Galactic TeV sources, dedicated multi-wavelength investigations for all known systems ($\sim10$ TGBs) have been performed with a number of X-ray, GeV, and TeV observations \citep[e.g.,][]{Takahashi2009,Corbet2012,Adams2021}. These studies uncovered diverse physical phenomena, including strong orbit-to-orbit variability \citep[e.g.,][]{Tokayer2021}, correlations between $\Gamma_X$ and $F_X$ \citep[e.g.,][]{BoschRamon2005, An2013}, variable $N_{\rm H}$ \citep[e.g.,][]{Malyshev2019,Tokayer2021}, and X-ray flares in certain  sources \citep[e.g.,][]{Chernyakova2015}. These findings highlight the complicated interactions between the particle flow, the companion's wind, disk \citep[e.g.,][]{Kefala2023}, and particle evolution in the IBS. The hard X-ray data from \nustar\ have added important insights into TGBs, e.g., possible magnetar-like X-ray pulsations in LS~5039 \citep[][]{Yoneda2020} \citep[although claimed to be spurious later; see ][]{Volkov2021} and the extension of  simple power-law emission to 20--30\,keV \citep[e.g.,][]{Tokayer2021,An2015} suggesting that particle cooling is not severe. Although these previous X-ray and gamma-ray studies revealed the complex emission mechanisms of the TGBs \citep[e.g.,][]{Dubus2015}, our understanding of the diverse phenomena specific to TGBs remains incomplete. More precise characterization of X-ray spectral variability ($\Gamma_X$, $F_X$, and $N_{\rm H}$), achievable by HEX-P, thanks to its contemporaneous observations in the broad 0.2--80\,keV band enabled by LET+HET, will help in discerning variabilities caused by the injection and cooling of relativistic particles and their interactions with the environment. 

The future prospects for TGB science are promising with the advent of CTAO and HEX-P. CTAO is expected to discover more TGBs \citep{Dubus2017} and measure the TeV spectral variations of  known TGBs on timescales as short as $\sim30$ min \citep{Chernyakova2019}. A larger sample size of TGBs will expand the parameter space and facilitate a deeper understanding of their diverse nature. 
Dedicated observation programs with HEX-P and CTAO will by far surpass the current studies involving \nustar, VERITAS, and H.E.S.S. Overall, 
 HEX-P is poised to make significant contributions to TGB astrophysics, enhancing our knowledge of these enigmatic binary systems and elucidating their complex emission mechanisms associated with relativistic shocks and jets in TGBs. 

%



\begin{table*}
\renewcommand\thetable{2}
{\small
\begin{center}
\caption{Potential HEX-P survey programs}
\begin{tabular}{lcccc}
\hline\hline
Program Description & Sources & Comments \\ 
\hline\hline
Galactic PeVatrons &  TBD (e.g., Dark PeVtrons) &  CTAO, LHAASO, HAWC, SWGO  \\
Hadronic PeVatrons & IceCube neutrino sources & IceCube  \\
Leptonic PeVatrons & PWNe detected above 100 TeV & CTAO, LHAASO, HAWC, SWGO \\
Young SNRs (X-ray variability) & Cas A, Tycho, G1.9+0.3 & Multi-epoch observations \\
Young SNRs ($^{44}$Ti science)  &  Cas A, Tycho, G1.9+0.3, G350.1-0.1 &  Combine multi-epoch observations$^*$\\
Sgr A* monitoring & Supermassive BH & EHT, GRAVITY, CTAO \\ 
Galactic NSM candidates & TBD  & COSI \\ 
\hline
\label{tab:survey_ideas}
\end{tabular}
\end{center}
}
Note: These are potential HEX-P survey ideas other than the primary science program listed in Table \ref{tab:obs}.  \\
$^*$ $^{44}$Ti science program will benefit from combining HEX-P data obtained from multi-epoch observations, which are intended for detecting X-ray variabilities from young SNRs.  
\end{table*}

\section{Conclusions} 

As our simulations demonstrate, HEX-P is poised to revolutionize X-ray views of Galactic particle accelerators, unraveling the origin of CRs up to the knee and beyond. Along with CTAO, HEX-P  will play a crucial role in identifying numerous  PeVatron candidates and their acceleration mechanisms. An extensive HEX-P survey of various types and stages of the particle accelerators associated with known SNRs, PWNe, star clusters, binaries, and BHs, will provide a broad picture of how particle acceleration, propagation, and cooling operate in different sources and environments.  
Finally, HEX-P will foster multi-messenger observation programs with other future missions such as CTAO, IceCube gen2, and COSI (see Table \ref{tab:survey_ideas} for potential HEX-P survey programs). HEX-P clearly stands out as the foremost X-ray observatory for particle acceleration astrophysics in the 2030s.

\section*{Author Contributions}

\S 1 (Reynolds).  \S 2 (Reynolds, Mori). \S 3 (Madsen and Garcia). \S 4 (Mori). \S 5 (Mori and Woo).  \S 6 (Bamba, Mori, Krivonos, Abdelmaguid). \S 7 (Tsuji, Mori, Park). \S 8 (An). \S 9 (Mori). 

Zoglauer (COSI). An, Eagle (Fermi). Grefenstette, Madsen, Mori, Woo, Zhang (NuSTAR). Bamba, Terada (XRISM). Nynka (Chandra). Younes (NICER). Bangale, Kumar, Shang, Woo (VERITAS).  



\section*{Acknowledgments} 
We are grateful to J. Wilms, T. Dauser, C. Kirsch, M. Lorenz, L. Dauner, and the SIXTE development team for their assistance with SIXTE simulations.
HA acknowledges support from the National Research Foundation of Korea (NRF) grant funded by the Korean Government (MSIT) (NRF-2023R1A2C1002718).
This work was financially supported by Japan Society for the Promotion of Science Grants-in-Aid for Scientific Research (KAKENHI) Grant Numbers, JP23H01211 (AB),
22K14064 (NT), 20K04009 (YT).
In Section~\ref{sec:ss433} (SS 433/W50 lobes), we thank Takahiro Sudoh and Kazuho Kayama for providing their results and Samar Safi-Harb for the image. 
We thank Ke Fang, Lu Lu, and Kelly Malone for their helpful discussion.



\bibliographystyle{Frontiers-Harvard} 


\bibliography{reference}

\end{document}